\documentclass[fleqn,usenatbib]{mnras}

\usepackage{newtxtext,newtxmath}

\usepackage[T1]{fontenc}

\DeclareRobustCommand{\VAN}[3]{#2}
\let\VANthebibliography\thebibliography
\def\thebibliography{\DeclareRobustCommand{\VAN}[3]{##3}\VANthebibliography}

\usepackage{graphicx}	
\usepackage{amsmath}	

\title[Atmosphere of GJ 3470\,b in high res]{
Constraints on atmospheric water abundance and cloud deck pressure in the warm Neptune GJ 3470\,b via CARMENES transmission spectroscopy}

\author[S. Dash et al.]{
Spandan Dash,$^{1,2}$\thanks{E-mail: Spandan.Dash@warwick.ac.uk}
Matteo Brogi,$^{3}$
Siddharth Gandhi,$^{1,2,4}$
Marina Lafarga,$^{1,2}$
Annabella Meech,$^{5}$
\newauthor
Aaron Bello-Arufe$^{6}$
and Peter J.\ Wheatley$^{1,2}$
\\
$^{1}$Department of Physics, University of Warwick, Coventry CV4 7AL, United Kingdom\\
$^{2}$Centre for Exoplanets and Habitability, University of Warwick, Coventry, CV4 7AL, United Kingdom\\
$^{3}$Department of Physics, University of Turin, Via Pietro Giuria 1, I-10125, Turin, Italy\\
$^{4}$Leiden Observatory, Leiden University, Postbus 9513, 2300 RA Leiden, The Netherlands\\
$^{5}$Department of Physics, University of Oxford, Keble Road, Oxford, OX1 3RH, United Kingdom\\
$^{6}$Jet Propulsion Laboratory, California Institute of Technology, Pasadena, CA 91109, USA
}

\date{Accepted XXX. Received YYY; in original form ZZZ}

\pubyear{2024}

\begin{document}
\label{firstpage}
\pagerange{\pageref{firstpage}--\pageref{lastpage}}
\maketitle

\begin{abstract}
Observations of cooler atmospheres of super-Earths and Neptune sized objects often show flat transmission spectra. 
The most likely cause of this trend is the presence of aerosols (i.e. clouds and hazes) in the atmospheres of such objects. 
High-resolution spectroscopy provides an opportunity to test this hypothesis by targeting molecular species whose spectral line cores extend above the level of such opaque decks. In this work, we analyse high-resolution infrared observations of the warm Neptune GJ 3470\,b taken over two transits using CARMENES (R $\sim$ 80,000) and look for signatures of H$_2$O (previously detected using HST WFC3+Spitzer observations) in these transits with a custom pipeline fully accounting for the effects of data cleaning on any potential exoplanet signal. We find that our data are potentially able to weakly detect ($\sim3\sigma$) an injected signal equivalent to the best-fit model from previous HST WFC3+Spitzer observations. However, we do not make a significant detection using the actual observations. Using a Bayesian framework to simultaneously constrain the H$_2$O Volume Mixing Ratio (VMR) and the cloud top pressure level, we select a family of models compatible with the non detection. These are either very high VMR, cloud-free models, solar-abundance models with a high cloud deck, or sub-solar abundance models with a moderate cloud deck. This is a broader range compared to published results from low-resolution spectroscopy, but is also compatible with them at a 1$\sigma$ level.
\end{abstract}

\begin{keywords}
exoplanets -- planets and satellites: atmospheres
\end{keywords}



\section{Introduction}\label{introduction}
High Resolution\footnote{At a resolving power $R = \lambda/\Delta\lambda \ge 25000$} Cross-Correlation Spectroscopy (HRCCS) is a method of characterising exoplanet atmospheres in transmission and emission at optical and infra-red wavelengths that has come of age in the last decade. The research focus has moved on from detections of single species \citep{snellen2010orbital, brogi2012signature, de2013detection, birkby2013detection, lockwood2014near} to detections of multiple species at the same time \citep{hawker2018evidence, cabot2019robustness, giacobbe2021five} to now finally constructing and using retrieval frameworks \citep{brogi2019retrieving, gibson2020detection}. These retrieval frameworks are capable of quantitatively estimating atmospheric parameters such as species abundances by taking into account biases arising from the removal of telluric contaminants in the Earth's atmosphere that are specific to HRCCS. Through a Likelihood framework, they also facilitate linking of high resolution spectroscopy to low resolution spectroscopy from space based telescopes such as HST, Spitzer and JWST, thereby providing a way to  characterise exoplanet atmospheres to greater precision than possible with either approach alone \citep{gandhi2019hydra, line2021wasp77Ab}.
\\
\\
Most of the above mentioned advances have been limited to working with hot and ultra-hot Jupiters. These objects, however, are not the most common outcome of planet formation. The most common are either planets with predominantly rocky cores, or planets (smaller than gas giants) with gaseous envelopes of H$_2$/He \citep{fulton2017california, owen2017evaporation, ginzburg2018core}. These two populations are separated by a radius valley \citep[as indicated by][]{fulton2017california}, which is a paucity of exoplanets at 1.5--2R$_{\oplus}$ separating a bimodal distribution, with peaks at 1.3 and 2.4R$_{\oplus}$, for observed exoplanets within a 100-day orbital period. For exoplanets falling in the 2--4R$_{\oplus}$ range i.e.\ above the radius valley as defined above and hence comprising the dominant population of exoplanets with atmospheres, there exists a compositional degeneracy between possible exoplanet varieties. It is difficult to differentiate between planets with icy water rich interior along with a thick H$_2$ rich atmosphere (called Water worlds, Ocean worlds or Hycean Worlds) from planets with a massive rocky iron rich core with a thin atmosphere just on the basis of their calculated densities from mass and radius parameters \citep{adams2008ocean, madhusudhan2021habitability, kempton2023water}. Precise atmospheric characterisation of such objects has been proposed as a solution that could break this compositional degeneracy \citep{adams2008ocean}. However, this may prove challenging at low spectral resolution because the observations of atmospheres of such smaller and cooler exoplanets often result in flat spectra in transmission, which is hypothesised to be due to the presence of aerosols (i.e. clouds and hazes) that act as opaque decks at specific altitudes and serve to flatten the spectra at such resolution \citep{kreidberg2014clouds, knutson2014featureless, crossfield2017trends}.
In contrast, HRCCS is potentially sensitive to molecular species whose line cores extend above the cloud deck in these objects \citep{de2014identifying, kempton2014high, gandhi2020seeing, hood2020prospects} and hence provide the necessary solution to break the compositional degeneracy among this exoplanet population. In addition to removing the compositional degeneracy, precise atmospheric characterisation can also help in understanding and tracing the formation histories of such objects, something which has already been attempted for hot Jupiters at high spectral resolution \citep{giacobbe2021five, line2021wasp77Ab, gandhi2023retrieval}. Studies at the same level of detail for smaller and cooler exoplanets have not been attempted yet, despite being theoretically feasible in the near future \citep{bean2021nature, bitsch2021dry}.
\\
\\
Considering the 
push towards atomic and molecular detections in cooler sub-Jupiters, attempts at characterisation of such candidates in the optical and infra-red have already started. There have been species detections and evidence of disequilibrium chemistry from warm giants like WASP 80\,b \citep{carleo2022gaps} and cool puffy Saturns like WASP 69\,b \citep{guilluy2022gaps}. However, attempts at detection of species have proved difficult so far, for instance in the cases of Super-Earths like GJ 486\,b \citep{ridden2022high} and 55 Cnc\,e \citep{deibert2021near, keles2022pepsi} and sub-Neptunes like WASP 166\,b \citep{lafarga2023hot}. As pointed out before, while HRCCS has the potential to look above the cloud deck in such objects, there are often various other factors which make this difficult. One factor is of course that reduced transit depths and shorter transit times (compared to hot Jupiters) together make it challenging to capture information about species in a single night of observation, which can instead be sufficient for hot and ultra-Hot Jupiters. Another factor is the accuracy of the model being correlated against. Hot and ultra-hot Jupiters' atmospheres being in thermochemical equilibrium in the deeper and mid-atmospheres with deviations in the upper parts primarily due to photochemistry makes it easy to generate models to correlate against. \citet{guilluy2022gaps} found that this will probably not hold true in cases of cooler and smaller exoplanets where the chemical diversity and inclusion of possible disequilibrium chemistry makes template model generation a complicated and time consuming process (also see \citealt{moses2013compositional}).
\\
\\
The idea that HRCCS could be leveraged to detect molecules above a cloud layer was already hypothesised by \citet{de2014identifying} and \citet{kempton2014high}. \citet{gandhi2020seeing} and \citet{hood2020prospects} were, however, the first authors to provide a quantitative estimate of the observational standards necessary to make a successful detection. Through simulations, \citet{gandhi2020seeing} found that the best fit cloudy model spectrum to HST WFC3+Spitzer observations in \citet{benneke2019sub} for detection of H$_2$O in the atmosphere of GJ 3470\,b would require at least 4 nights ($\sim$8 hours) of observations (in total) to be unambiguously detected using any of CARMENES, GIANO or SPIRou, which are among the best performing spectrographs in use today. A similar conclusion was also reached by \citet{hood2020prospects} who found that at least 10 hours of observations (in total) in the K and H bands would be necessary to detect CO and H$_2$O for a photochemical haze model that matched GJ1214\,b's HST observations \citep{kreidberg2014clouds}. Hence, getting good quality nights from a combination of spectrographs covering at least the J, K and H bands is of great importance if detections of molecules using HRCCS is to be extended to the sub-Neptune/super-Earth regime. Even in the case of observations from multiple instruments being available, a mechanism has to exist in order to be able to combine the analyses from the nights of observations from different instruments with varying resolutions. The Cross-Correlation Function to log-likelihood framework (CCF-to-$\log(L)$) from \citet{brogi2019retrieving} and \citet{gibson2020detection} both enable that approach. 
These authors also showed that traditional HRCCS detrending algorithms leave an impression on any exoplanet signal in the data, which necessitates accounting for those specific biases before analysing any retrieved results. This process, sometimes called \emph{Model Reprocessing}, involves passing the model template spectrum through the exact same detrending procedure that is applied to the raw data before computing a Likelihood value. Likelihood values allow the use of Bayesian statistics to compare complex atmospheric models. This approach, when coupled with an MCMC algorithm, results in proper retrievals of planetary parameters such as the chemical abundances of detected elements. The implementation of Bayesian analysis has thus allowed HRCCS to move beyond the realm of mere detections of species to proper exoplanetary characterisation at the level of results reported from low resolution studies like HST and now JWST \citep[compare e.g.][for the case of WASP-77\,Ab]{august2023confirmation, line2021wasp77Ab}. The exact way of implementing model reprocessing has varied across works but the reasoning remains that using this procedure is necessary for proper quantification of any retrieved results from the HRCCS analysis. 
\\
\begin{table*}
    \centering
    \begin{tabular}{|c|c|c|c|}
    \hline
    \textbf{Parameter} & \textbf{Description} & \textbf{Value} & \textbf{Reference} \\
    \hline
    \hline
    M$_{\star}$ & Stellar mass & $0.51^{+0.06}_{-0.06}$ M$_{\odot}$ & \citealt{kosiarek2019bright} \\
    R$_{\star}$ & Stellar radius & $0.48^{+0.04}_{-0.04}$ $R_{\odot}$ & \citealt{kosiarek2019bright} \\
    \hline
    M$_\mathrm{P}$ & Planet Mass & $12.58^{+1.31}_{-1.28}$ M$_{\oplus}$ & \citealt{kosiarek2019bright} \\
    R$_\mathrm{P}$ & Planet Radius & $3.88^{+0.32}_{-0.32}$ R$_{\oplus}$ & \citealt{kosiarek2019bright} \\
    $a$ & Semi-major axis & $12.92^{+0.72}_{-0.62}$ $R_{\star}$ & \citealt{kosiarek2019bright} \\
    T$_\mathrm{eq}$ & Equilibrium Temperature & $615^{+16}_{-16}$ K &  \citealt{bonfils2012hot,kosiarek2019bright} \\
    $P_\mathrm{P}$ & Orbital Period & $3.336649^{+0.000084}_{-0.000081}$ days & \citealt{kosiarek2019bright} \\
    T$_\mathrm{0}$ & Transit Mid-Point & $2456340.725595^{+0.00010}_{-0.00010}$ BJD & \citealt{stefansson2022warm}\\
    T$_{14}$ & Transit Duration & $1.918^{+0.024}_{-0.024}$ hours & \citealt{kosiarek2019bright} \\
    $e$ & Eccentricity & $0.114^{+0.052}_{-0.051}$ & \citealt{kosiarek2019bright} \\
    $v_\mathrm{sys}$ & Systemic Velocity & $26.09^{+0.25}_{-0.25}$ km\,s$^{-1}$ & \citealt{brown2018gaia} \\
    $i$ & Orbital Inclination & 89.13$^{\circ +0.26^{\circ}}_{-0.34^{\circ}}$ & \citealt{awiphan2016transit} \\
    $K_\mathrm{P, cir}$ & Exoplanet Semi-amplitude velocity (assuming circular orbit) & $94.1^{+8.6}_{-8.6}$ km\,s$^{-1}$ & Expected Value, this study \\
    $K_\mathrm{P, ecc}$ & Exoplanet Semi-amplitude velocity (assuming eccentric orbit) & $94.7^{+8.6}_{-8.6}$ km\,s$^{-1}$ & Expected value, this study \\
    \hline
    \end{tabular}
    \caption{Parameters of the GJ 3470 planetary system used in this study and their values}
    \label{tab:system_par}
\end{table*}
\\
With the above approach in mind, in this work we modify existing HRCCS algorithms to treat spectral sequences with relatively low S/N and where only a small fraction of the spectra correspond to the exoplanet transit. We then use this revised pipeline to try and detect H$_2$O in the infra-red transmission spectrum of the sub-Neptune GJ 3470\,b using 2 nights of high resolution data sets from CARMENES \citep{quirrenbach2016carmenes}. This sub-Neptune of mass of $12.58^{+1.31}_{-1.28}$ M$_{\oplus}$ and radius of $3.88^{+0.32}_{-0.32}$ R$_{\oplus}$ orbits an M-dwarf star of mass $0.51^{+0.06}_{-0.06}$ M$_{\odot}$ with an orbital period of $3.336649^{+0.000084}_{-0.000081}$ days and a short transit duration of $1.918^{+0.024}_{-0.024}$ hours (all values from \citealt{kosiarek2019bright}, for convenience, all stellar and planetary parameters used in this paper are provided in Table \ref{tab:system_par}). It has an equilibrium temperature of $615^{+16}_{-16}$ K assuming a Bond Albedo value of 0 \citep{bonfils2012hot,kosiarek2019bright}, but the exoplanet is inflated with a neutral hydrogen atmosphere filling its Roche lobe \citep{bourrier2018hubble} as well as an escaping He outflow \citep{palle2020he, ninan2020evidence}. Low resolution results from HST and Spitzer presented by \citet{benneke2019sub} posit that the exoplanet has a hydrogen dominated atmosphere 
with a $>$5$\sigma$ detection of H$_2$O at 1.4 $\mu$m and a substantial methane depletion. In addition, they also found evidence of high opacity and high altitude clouds which resulted in attenuation of features, together with a significant drop off in opacity at 2-3$\mu$m possibly due to Mie scattering, which provided an estimate of the grain size of the clouds. As mentioned above, \citet{gandhi2020seeing} have determined that their high resolution best fit model template could be detected using 4 nights of observations. Hence, our study here provides an experimental continuation of their work using our new HRCCS pipeline and describes the possible challenges we face in doing so. Section \ref{method} of the paper describes the construction of our pipeline, the way we process our model and also how we incorporate the \emph{Likelihood} framework approach to try and detect a template forward model as well as to do a model selection on a grid of models. In Section \ref{results}, we look at results from using our method to retrieve an injected model as well as from application of the pipeline on 2 nights of observed data from CARMENES. In Section \ref{discussion} we discuss the possible pitfalls, limits to assumptions, further applications and the future outlook of the procedure before we conclude in Section \ref{conclusion}.

\section{Methodology}\label{method}

\subsection{Observations and pre-processing}\label{21}
\begin{figure}
    \centering
    \includegraphics[width = \columnwidth]{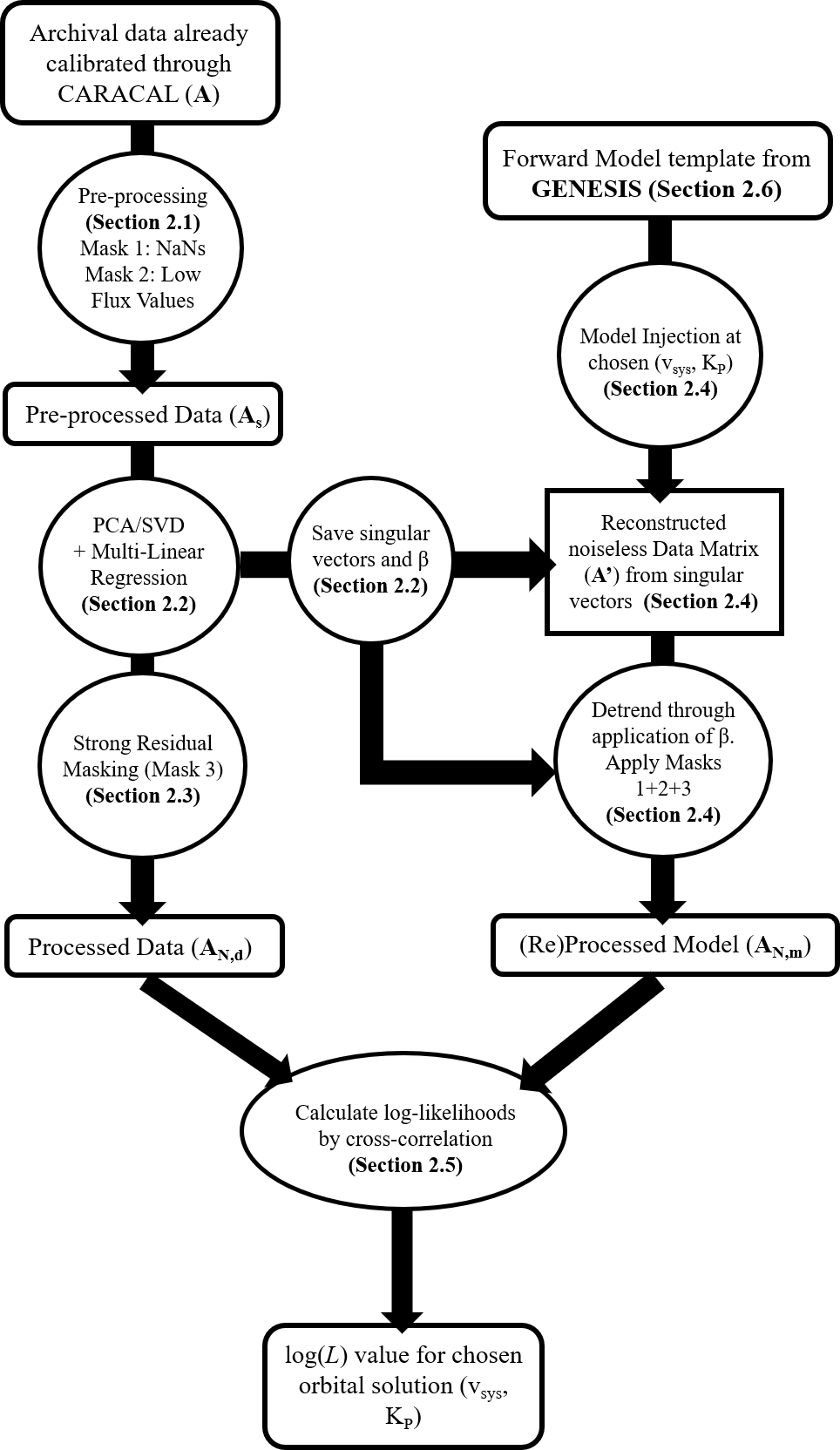}
    \caption{A schematic representation of the procedures covered in Sections \ref{21} to \ref{26} for a chosen orbital solution, with the notations for the matrices and variables being the same as in the main text in each section.}
    \label{fig:methodology}
\end{figure}

In this study, we use two nights of publicly available archival observations of the GJ 3470 system taken using the CARMENES instrument mounted on the 3.5m telescope at the Calar Alto Observatory. CARMENES \citep{quirrenbach2016carmenes} consists of two high resolution spectrographs covering 0.52-0.96 $\mu$m (VIS channel, R = 93,400) and 0.96-1.71 $\mu$m (NIR channel, R = 81,800). Each channel collects light from two fibres (A and B). For our analysis, we only use data generated from the NIR channel using fibre A placed on the target, which is obtained from cross-dispersed light (from an echelle grating) falling across 28 orders on two 2048$\times$2048 pixel Hawaii-2RG CCDs. Fibre B is pointed either towards a Fabry-Perot etalon or the sky in case of faint targets, as no nodding is typically applied \citep{caballero2016carmenes}. CARMENES is an ultra-stable spectrograph that is designed to detect low mass planets around M-dwarf stars using accurate radial velocity measurements on the order of a few m\,s$^{-1}$ on a long term basis. Its stability ensures that there are no significant shifts in wavelength for the absorption lines due to H$_2$O present in the Earth's atmosphere (telluric contamination) during the course of our observations. Hence, we do not need any additional pixel-wavelength solution correction for our data and depend on the wavelength solutions already generated by the CARMENES pipeline (CARACAL) by wavelength calibration through a Fabry-Perot etalon \citep{bauer2015calibrating}. The pixel-wavelength solution is also the same for each spectrum taken during the night and is calculated in the rest frame of the observatory. 
\\
\\
The first observation was taken on the night of 26 and 27 December 2018 with 34 exposures covering an orbital phase range from -0.032 to +0.036 (with 0 indicating mid-transit phase), with airmass values between 1.91 and 1.14 for the corresponding phases, and the ambient relative humidity varying between 20\% and 24\% throughout the observation. The integration time for each exposure during this observation was 500s. The second observation was taken on the night of 5 and 6 January 2019 with 35 exposures covering an orbital phase range from -0.032 to +0.036, with the airmass varying between 1.48 and 1.25, and the ambient relative humidity varying between 19\% and 28\% throughout the observation. The integration time for each exposure for the second night was the same as the first night. There is a third, more recent transit of GJ 3470\,b available in the CARMENES archive, but it seems to have been taken with a different observing strategy incompatible with the automated CARMENES pipeline. Hence, we have not used that night in this analysis. With the planetary transit phases lying between -0.012 and +0.012, all these observations include exposures that cover the entirety of the transit (extending through 12 exposures in each night, which is about 1/3rd of the total exposures in each night), with a substantial number of out-of-transit exposures as well. We note here that the flux value in the exposure files is not the raw flux value obtained from observation of the target, but a value obtained by optimal extraction after dividing through the spectrum of a calibration lamp. Such extraction is the standard procedure applied by the CARACAL pipeline \citep{zechmeister2014flat, caballero2016carmenes}. As such, the spectra are corrected for the blaze function, but are also imprinted by the spectrum of the calibration lamp (i.e. a broad-band slope).
\\
\\
Data obtained from the NIR channel (and limited to fibre A) of CARMENES are in the form of one FITS file for each exposure. From all the FITS files of an observing night, we create cuboids of flux (this cuboid is denoted henceforth as $\textbf{A}$), flux error (cuboid denoted as \textbf{$\varepsilon$}) and wavelength. These cuboids have dimensions $( n_\mathrm{orders} \times  n_\mathrm{spectra} \times n_\mathrm{pixels} )$, where $n_\mathrm{orders} = 28$ denotes the number of orders 
, $n_\mathrm{spectra}$ the number of exposures (33 and 34 for the first and second transits, respectively), and $n_\mathrm{pixels} = 4028$ the number of pixels, each with an associated wavelength solution\footnote{Dimensions of Orders and Spectra can be interchanged as long as the way we do the detrending remains the same. We use this convention to maintain consistency with previous works.}. Each order-wise slice of a flux cuboid looks like the one shown in Panel (a) of Figure \ref{fig:steps}. There is a variation of flux across rows and columns, plus many 
dark vertical features which are caused by telluric absorption in the Earth's atmosphere. From this point on, 
the rest of the analysis is performed order by order, unless specified otherwise. We provide a flow-chart representation of the processing steps 
covered in Sections \ref{21} to \ref{26} in Figure \ref{fig:methodology}.
\\
\\
Pre-processing of the flux cuboid involves masking NaN values (a byproduct of the aforementioned pipeline normalisation) to 0, followed by masking low flux values to 0 by using a threshold of 5 percent of the median of the highest 300 flux values in each order. This ``low-flux correction'' should result in the more strongly saturated absorption features in each order due to telluric absorption being masked. This masks around 6\% of the total flux cuboid for night 1 and 4.2\% of the total flux cuboid for night 2 with the majority of the masked values falling in Orders 8-10 and 18-22 (when we denote the first order as the 'zeroth' order). This is expected as these orders fall in the wavelength range where telluric absorption is known to be the highest. 
\\
\\
\subsection{Detrending  using Principal Component Analysis}\label{22}

Searching for a very faint changing exoplanet signal within the raw data cuboid (\textbf{A}) means to actually differentiate between the multiple contributions to flux variations that particular data cuboid would have, from both astrophysical and instrumental effects. So, it is not the actual absolute values of the fluxes we are concerned with, but rather the variation of flux with time, wavelength, airmass, detector response etc. As such we need to first \emph{equalise} contributions from any large scale features that might cause the majority of the variation and dominate the detrending process otherwise. 
One way to equalise the contributions from all sources of variations is to \textit{standardise} our data. We do this by first subtracting the mean from each pixel column (across time) and then dividing all mean subtracted values by the standard deviation of the corresponding pixel column. This equally weights the flux data over each pixel (i.e. each wavelength). The standardised matrix at the end of this step is then denoted as $\textbf{A}_{\textbf{S}}$. Previous works standardised their data on an order basis by comparing the spectrum obtained from each row to the median spectrum of that order. Either of these methods should work, but only one step of standardisation needs to be applied before further detrending.
\\
\\
The pre-processed and standardised data cuboid thus obtained has variations from multiple sources (such as telluric absorption, barycentric motion of the observer, motion of the star around the common centre of mass of the exoplanet system) but we only need 
to look for variations that correspond to an exoplanet signal. Ideally, for an ultra-stable instrument like CARMENES, only the exoplanet signal will clearly shift in wavelength over time on the order of some km\,s$^{-1}$ and all other contributions will remain static or experience small shifts in wavelength, much less than the velocity resolution limit of the instrument. Hence, any kind of processing that can remove these unchanging/very slowly shifting contributions, which also make up for the bulk of the variations in the standardised flux cuboid, is sufficient to our purposes. Many approaches have been used for this step, 
including methods such as fitting a low order power of the geometric airmass to the flux in each spectral channel (known as the airmass method) \citep{brogi2012signature}, through Singular Value Decomposition (SVD) as a more general use of a Principal Component Analysis (PCA) technique \citep{de2013detection}, using SYSREM (a variant of the PCA based approach) \citep{birkby2013detection, tamuz2005correcting, mazeh2007asp} and through direct modelling of telluric features for removal \citep{lockwood2014near}. Each of these detrending techniques have been used to yield detections of atoms and molecules, with the unsupervised SVD/PCA based approaches being shown as better performing \citep{cabot2019robustness}. However, optimising detrending on an order by order basis using any of the algorithms has been shown to have the potential to bias detection of an artificially introduced exoplanet signal of the same strength as an expected signal in the data, compared to the case of applying the same algorithm uniformly across all orders \citep{cheverall2023robustness}. Keeping that pitfall in mind, in this work we use the simpler unsupervised SVD approach keeping the processing steps same for each order (including the number of SVD/PCA components used for detrending). This also makes it easier for us to be in control at every stage, and to understand and correct any potential problems that might emerge.
\\
\\
The SVD approach decomposes the standardised data matrix ($\textbf{A}_{\textbf{S}}$) with dimensions ($n_\mathrm{orders}$ $\times$ $n_\mathrm{spectra}$ $\times$ $n_\mathrm{pixels}$) into a product of three matrices:
\begin{equation}
    \textbf{A}_{\textbf{S}} = \textbf{U}\textbf{S}\textbf{V}^{\mathrm{T}}.
\end{equation}
Hence, effectively it decomposes each of the $n_\mathrm{orders}$ 2D matrices of size ($n_\mathrm{spectra}$ $\times$ $n_\mathrm{pixels}$) into a matrix of left singular vectors ($\textbf{U}$) with shape ($n_\mathrm{spectra}$ $\times$ $n_\mathrm{spectra}$), a diagonal matrix vector ($\textbf{S}$) of shape ($n_\mathrm{spectra}$ $\times$ $n_\mathrm{pixels}$) and a matrix of right singular vectors ($\textbf{V}$) of shape ($n_\mathrm{pixels}$ $\times$ $n_\mathrm{pixels}$). $\textbf{V}^{\mathrm{T}}$ in the equation above is the transpose of \textbf{V}. We would like to note that this way of doing SVD where the first dimension of the 2D standardised flux matrix to be decomposed corresponds to the dimension of time is called time-domain SVD, and the singular vector matrices obtained are in the time domain. Henceforth in this paper, we just write SVD to denote time-domain SVD. A variant of this approach swaps the time and wavelength dimensions before standardisation and and then performs the SVD, and is called hence wavelength-domain SVD.
\\
\\
To save computational time and data, a variant of the SVD approach, called Reduced SVD, cuts down the shapes of matrices $\textbf{U}$, $\textbf{S}$ and $\textbf{V}$ to ($n_\mathrm{spectra}$ $\times$ $r$), ($r$ $\times$ $r$) and ($n_\mathrm{pixels}$ $\times$ $r$) respectively, where $r$ is \textbf{min}($n_\mathrm{spectra}$, $n_\mathrm{pixels}$). This is similar to Step 3 of the data reduction process in \citet{pino2022gaps}. Generally for CARMENES data for transmission spectra, the number of spectra is vastly lower than the number of pixels (4080) and hence $r$ is always equal to $n_\mathrm{spectra}$. This places a fundamental limit on the number of singular vectors each order is decomposed into and on how much flux variation is packaged into each of those vectors. 
\\
\\
One of the features of SVD is that the higher the singular value for a corresponding singular vector, the higher its contribution to flux variations. The diagonal matrix obtained by  SVD\footnote{numpy.linalg.svd(order\_matrix, full\_matrices = False) for reduced SVD} provides all singular values ranked from highest to lowest. Now that we have the singular values and singular vectors, we could reconstruct the original data matrix $\textbf{A}$ exactly by linearly combining all the singular vectors and values. Alternatively, we could only select the first $k$ eigenvectors, and reconstruct a partial representation of $\textbf{A}$, where most of the time-varying nuisance components are captured while the exoplanet signal is hopefully left out. However, we performed our SVD on the standardised matrix $\textbf{A}_{\textbf{S}}$ and not the pre-processed matrix $\textbf{A}$. To account for this fact, we need to refit the singular values 
accordingly. To do this, we perform a multilinear regression by selecting for the first $k$ singular vectors (plus an additional vector of ones to properly bias for the fact that we standardised the data matrix during the SVD process) to reconstruct a noiseless data matrix. This output $\textbf{A'}$ as the best fit matrix computed by multilinear regression is calculated as:
\begin{equation}
    \textbf{A'} = \textbf{U}(\textbf{U}^\mathrm{T}\textbf{U})^{-1}\textbf{U}^\mathrm{T}\textbf{A}.
\end{equation}
We henceforth denote $\textbf{U}(\textbf{U}^\mathrm{T}\textbf{U})^{-1}\textbf{U}^\mathrm{T}$ as $\beta$. This simplifies equation (2) to:
\begin{equation}
    \textbf{A'} = \beta\textbf{A}.
\end{equation}
Hence, the action of $\beta$ can be seen as the functional effect of processing/detrending (through a matrix multiplication) on the original data matrix $\textbf{A}$ resulting in a noiseless output matrix $\textbf{A'}$. This noiseless data matrix is then used to divide the processed data matrix to get a normalised data matrix $\textbf{A}_{\textbf{N}}$:
\begin{equation}
    \textbf{A}_{\textbf{N}} = \textbf{A} / \textbf{A'}.
\end{equation}
$\textbf{A}_{\textbf{N}}$ is effectively the \emph{matrix of residuals} left over after the detrending process, now clustered around 1.0 being the product of a normalisation. $\textbf{A'}$ and $\beta$ are both saved to reproduce the effects of this analysis on each forward model (see Section \ref{24}) as also done in \citet{gibson2022relative} for the case of SYSREM.
\\
\\
Typically to have finer control over how much variation we remove with each vector, the total number of singular vectors should be large. However, as pointed above, this is restricted by the value of $r$, which in our case is the total number of spectra/exposures taken in the observation. The nights of observations we use for this work have 33 and 34 spectra respectively. Since we ultimately 
only use the spectra (in transit) containing the exoplanet signal to avoid any other potential contamination, a case might be made for discarding all the out-of-transit spectra 
as early as possible in the analysis, e.g. prior to SVD/PCA. However, if we were to limit the detrending process to only the data in transit (which is about 1/3rd the size of the total observation), it would reduce the number of singular vectors accordingly and we would artificially force the SVD algorithm to capture most of the data variance within the first 1-2 components. This would result in imperfect telluric removal, because in this case we would approximate a combination of non-linear effects (astrophysical and instrumental) with a too small linear combination. Furthermore, adding subsequent components would cause the removal of the actual exoplanet signal within the first few singular vectors. 
\\
\\
We indeed tested the above, and found that using just the data in transit resulted in removal of a \emph{nominal} exoplanet signal\footnote{This is defined as an artificially introduced model signal into real data with its strength same as an expected exoplanet signal from GJ 3470\,b (also see Section \ref{24} for the process of introducing this signal)} equivalent to the model calculated in \citet{gandhi2020seeing}, within the first 2-3 singular vectors. This number went up to $>$ 7 if the entire sequence was used instead. There is also the added fact that the number of exposures should be large for robust strong residual masking (see Section \ref{23}). For an ideal case it would be better to have more singular vectors to work only using the data within the exoplanet transit phases. Hence, our recommendation for future observations is to obtain as many exposures as possible during the exoplanet transit, at least when the brightness of the target allows for sufficient signal to noise even with relatively short exposures. This issue with the number of collected spectra is not present in the case of emission spectroscopy where the exoplanet signal would ideally span the entire night of observation, except during secondary eclipse. Using the entire night for detrending in transmission spectroscopic analysis has a few caveats as we shall see in Section \ref{24}.

\subsection{Masking strong residuals}\label{23}
\begin{figure*}
    \centering
    \includegraphics[width = 2\columnwidth]{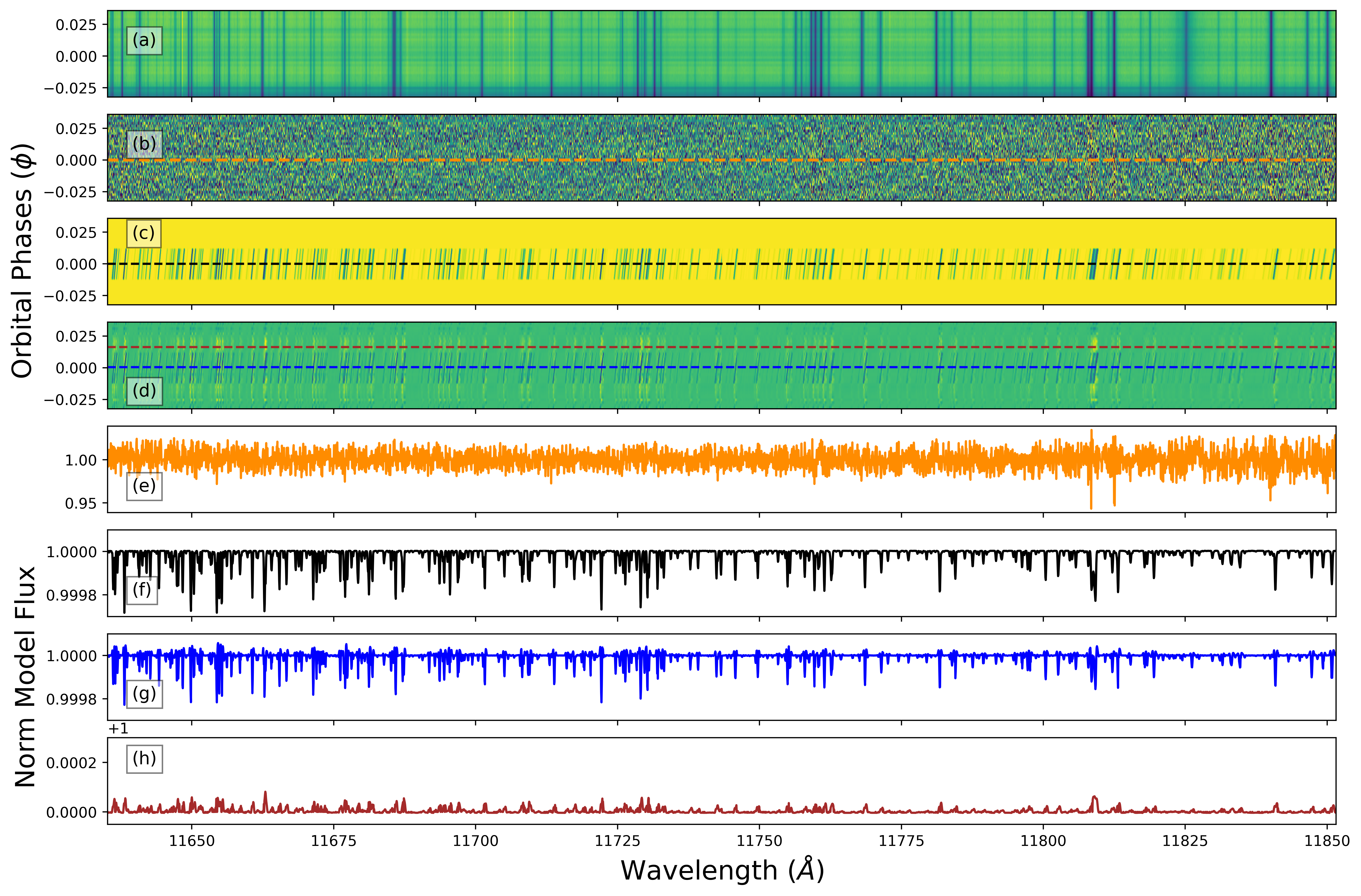}
    \caption{SVD/PCA based detrending process and the effect of model reprocessing for order number \textbf{11} of Night 1 of CARMENES observation. \textbf{(a)} Data after calibration through CARACAL, \textbf{(b)} After SVD/PCA based detrending and strong residual masking - the orange line shows the cross-section along which the spectra for (e) is drawn, \textbf{(c)} Nominal exoplanet model signal that is to be injected into noiseless data matrix before reprocessing - the black line represents the cross-section for drawing the spectrum in (f), \textbf{(d)} Reprocessed model signal showing the model signal + artefacts in the out-of-transit phases - blue is the cross-section for drawing (g) and reddish brown for (h), \textbf{(e)} Cross-section at mid-transit time (Phase $=$ 0) for the processed and normalised data in (b) showing variations around 1.0, \textbf{(f)} Same as (e) but for the nominal model signal in (c), \textbf{(g)} Same as (e) but for the reprocessed model signal in (d). The differences between (f) and (g) notably in the line depths and wing shapes illustrate the effect of the SVD/PCA based detrending procedure on the the model signal and by definition any actual signal that would be present in the data, \textbf{(h)} Cross-section at an out-of-transit time (Phase $\approx$ 0.02) showcasing the nature of the artefacts caused due to model reprocessing with a similar scale as in (g).}
    \label{fig:steps}
\end{figure*}
The deviations from 1.0 in the normalised residual matrix $\textbf{A}_{\textbf{N}}$ are due to multiplicative contributions\footnote{Variation in flux is assumed to follow the Beer-Lambert law where the variation caused due to absorption/emission by light passing through different media follows an exponential factor multiplying the original flux; hence we make sure to normalise anything by division of fluxes rather than subtraction. An alternate approach would be to work in logarithmic flux space (e.g. in magnitudes) in which case subtraction would be more appropriate. The original paper introducing the SVD approach in \citet{de2013detection} used subtraction.} to fluxes from other sources, including white (gaussian) noise, not accounted for by the considered singular vectors, also including the actual time-varying exoplanet signal. However, there can be unaccounted sources of variations like emission from the Earth's atmosphere, cosmic rays etc. To remove these, we undertake an additional strong residual masking step to mask the strongest residuals in $\textbf{A}_{\textbf{N}}$ after the SVD/PCA based detrending. Each CARMENES exposure comes with flux error values provided by the pipeline. Hence, as already mentioned in Section \ref{21} for fluxes, a flux error cuboid is also constructed for each night. From this cuboid, a relative flux error cuboid (\textbf{$\varepsilon_{r}$}) can be calculated by dividing the flux error cuboid by the flux cuboid i.e. \textbf{$\varepsilon_{r}$} $=$ \textbf{$\varepsilon/$}\textbf{A}. From the relative error cuboid, the relative RMS (Root Mean Square) error over each individual pixel column (i.e. each wavelength over time, with dimension $n_\mathrm{spectra}$) can be calculated order-wise, and gives an estimate of the limit of relative error variation for that pixel column (i.e. particular wavelength value) over that observational night. 
In other words, in the case of a perfect detrending and neglecting astrophysical noise, each normalised pixel of the  $\textbf{A}_{\textbf{N}}$ cuboid should be drawn at random from a Gaussian distribution with mean 1 and $\sigma$=RMS. Since there are $n$ $\leq$ $n_\mathrm{spectra}$ independent observations in each data column\footnote{Explicitly removing pixels previously flagged as NaN or low-flux}, each column should contain a sample of $n$ randomly drawn values, thus there is a probability $P = 1/n$ that one of these values will depart from 1 by more than $m$ times the RMS. The value of $m$ (the ``number of standard deviations") can be derived using a two tailed test. This is calculated via the Inverse Survival Function of a Normal distribution\footnote{implemented via \tt{scipy.stats.norm.isf($2/n$)}}, where $P$ is split equally between the positive and negative tail as both positive and negative deviations are acceptable. Any measured deviation exceeding $m\times$RMS is statistically unlikely to be drawn from a Gaussian distribution.
\\
\\
To implement the above, first 1.0 is subtracted from the detrended and normalised flux cuboid (\textbf{A$_\mathrm{\textbf{N}}$}, whose values vary around 1.0) order-wise to get a residual matrix \textbf{A$_\mathrm{\textbf{N},r}$}. The absolute variations in residuals above 0 over each pixel column in \textbf{A$_\mathrm{\textbf{N},r}$} are then compared to the threshold $m\times$RMS calculated in the preceding paragraph. Then the original pixels in the same pixel column in \textbf{A$_\mathrm{\textbf{N}}$} are masked if their corresponding residuals in \textbf{A$_\mathrm{\textbf{N},r}$} falls over the threshold. If 3 or more values over time for the same pixel column are masked in \textbf{A$_\mathrm{\textbf{N}}$} using the process above, then we mask the entire pixel column (i.e. over time) as it could have had instrumental problems over that observation and we want to be conservative in our approach to reduce any false positives. The processed and masked data cuboid that is left after this step is named as \textbf{A$_\mathrm{\textbf{N},d}$}. One order from the first night of observation using CARMENES at the end of the whole detrending and masking process (i.e. an order of \textbf{A$_\mathrm{\textbf{N},d}$}) looks like Panel (b) in Figure \ref{fig:steps}. Ideally the variation of values in this matrix should be distributed around 1.0 (similar to \textbf{A$_\mathrm{\textbf{N}}$}). This is indeed mostly the case as shown in a row-wise cross-section of an order of that matrix (orange dashed line in Panel (b) where the cross-section values are shown in Panel (e) in the same figure). The variations from 1.0 in Panel (e) follow that trend except for pixels where telluric contamination is the highest (e.g. the values at about 11810 \AA) because the relative errors are inevitably large at those values allowing for a slightly larger strong residual masking threshold calculated by the algorithm compared to the rest of the pixels (as seen in Panel (a)). 
\\
\\
The difficulty of detecting an exoplanet signal is made clear once one realises that any signal due to water absorption through a moving exoplanet signal (see Panel (c) in Figure \ref{fig:steps}) would fall close to these saturated telluric water features as well. In addition, unlike past studies which focused on ultra-hot and hot Jupiters, smaller and cooler exoplanets orbiting M dwarfs like GJ 3470\,b will have a smaller orbital velocity. Therefore the change in their Radial Velocity (RV) with time will be smaller, increasing the likelihood to be captured with the other stationary components by the SVD/PCA based processing algorithm. Hence, there is always the possibility that much of the signal is either lost or attenuated or simply distorted in the process of detrending itself.

\subsection{Model (Re)processing}\label{24}
\citet{brogi2019retrieving} first showed that application of a detrending algorithm, such as the one using SVD/PCA presented in Section \ref{22}, leaves an impression on the exoplanetary signal itself. This has led to studies finding different ways to process the forward model for every detrending algorithm used (see \citealt{gibson2022relative} for an example). Since the processing step is applied twice - once on the data and then on the model - this 
step is commonly referred to as \emph{Reprocessing}. However, most studies have remained limited to analysis of emission spectra, where the exoplanetary signal almost always extends over the entire night of observations, and for hot Jupiters. Analyses of transmission spectroscopic observations at high resolution have been so far scarce and only limited to ultra-hot Jupiters in the optical \citep{gibson2022relative, gandhi2023retrieval}. In this study, we aim to devise a processing algorithm which can work well on transmission spectra analysis in the infrared, where the signal (i) would only be present in a small part of the night (which would generally hold for smaller and cooler exoplanets like GJ 3470\,b under current observational strategies), (ii) is also characterised by molecules exhibiting a larger forest of lines (compared to the optical), and (iii) also falls close to saturated lines due to telluric contamination due to the Earth's atmosphere (as pointed out in the concluding paragraph of Section \ref{23}).
\\
\\
\citet{gibson2022relative} introduced a fast model reprocessing technique tuned for SYSREM. In this study, we extend their technique for our SVD/PCA-based approach. In equation 3, we saw that $\beta$ was effectively a function which transformed the original data matrix into a noiseless output matrix (through matrix multiplication). We use this to our advantage by saving it for use when we process the model. 
\\
\\
We also save the noiseless data matrix $\textbf{A'}$ into which we introduce the variations from any forward model we plan to cross-correlate with at the nominal level (a process called \emph{injection}). The injection step begins by first broadening it using a Gaussian kernel to the instrumental resolution (R) and then normalising its continuum through division by the maximum value of the continuum in the wavelength range considered. This is then followed by multiplying fluxes at any given orbital phases (spectra) in $\textbf{A'}$, by the additional absorption contributions of the model variations. These variations are calculated by Doppler shifting the broadened and continuum normalised model to wavelengths calculated using a chosen exoplanet orbital solution through spline interpolation. Such orbital solutions are characterised by two variables: $v_\mathrm{sys}$ (the systemic velocity) and $K_\mathrm{P}$ (the exoplanet semi-amplitude velocity) through the equation:
\begin{equation}\label{eq:full_rv}
    v_\mathrm{P}(t) = v_\mathrm{sys}-v_\mathrm{bary}(t)+K_\mathrm{P}\{ \cos[f(t)+\omega] + e\cos(\omega)\}.
\end{equation}
Here, $v_\mathrm{bary}(t)$ are the barycentric radial velocities of the observer throughout the observation, which are obtained from the same FITS files used to construct the data cuboid. $\omega$ is the longitude of periastron, $e$ is the eccentricity of the exoplanet's orbit and $f(t)$ is the true eccentric anomaly calculated as:
\begin{equation}
    f(t) = 2\arctan\bigg{(}\frac{\sqrt{1+e}}{\sqrt{1-e}}\bigg{)}\tan\bigg{(}\frac{E(t)}{2}\bigg{)},
\end{equation}
where $E(t)$ is the Eccentric anomaly which is obtained by solving the Kepler equation connecting it to the Mean anomaly ($M(t)$) numerically using the Newton-Raphson method:
\begin{equation}
    E(t) - e\sin[E(t)] - M(t) = 0.
\end{equation}
The time varying $M(t)$ is found through its relation to the exoplanet orbital phases:
\begin{equation}
    M(t) = 2\pi\phi(t) - \pi/2 -\omega.
\end{equation}
The exoplanet orbital phases ($\phi(t)$) are calculated from the Barycentric Julian Dates provided in the same FITS files from which we construct our data cuboid. The values of $e$ and $\omega$ for GJ 3470\,b are known to differ from a circular orbit \citep{stefansson2022warm} but we find that this leads to $<$1 km\,s$^{-1}$ increase of the measured planet $K_\mathrm{P}$ compared to assuming a circular solution (see Section \ref{25}). Hence, we just use a circular solution henceforth but still provide the entire workflow for general use. In the circular limit, Eq.~\ref{eq:full_rv} reduces to:
\begin{equation}
    v_\mathrm{P}(t) = v_\mathrm{sys}-v_\mathrm{bary}(t)+K_\mathrm{P}\{ \sin[2\pi\phi(t)]\},
\end{equation}
where all the symbols have the same meaning as in the previous equations.
\\
\\
After model injection into the reconstructed noiseless data matrix $\textbf{A'}$, this data cube will have variations accounted for by the number of singular vectors used to reconstruct the cube, plus an additional source of flux variation due to an injected exoplanet signal within the phases covering the transit. Initially, it might be assumed that the exoplanet signal is accounted for by an additional singular vector with its corresponding singular value. Processing this noiseless and injected data cube through the matrix multiplication of $\beta$ (calculated and saved while performing multilinear regression to get $\textbf{A'}$ from $\textbf{A}$, see Equation 3 in Section \ref{22}) should then ideally remove all of the flux variation due to the singular vectors accounted for in $\beta$ while leaving the excess injected model as an output. We then apply to the reprocessed model the exact same masks (masking NaN and low flux values while pre-processing in Section \ref{21}, and strong residual masking in Section \ref{23}) that were applied throughout the analysis of the observed data.
\\
\\
At the end of the process outlined above, we should ideally be left with only the injected model in the in-transit phase spectra (\textbf{A$_\mathrm{\textbf{N},m}$}). However, we see that there are still some small differences in normalised flux between in-transit spectra rows for some orders falling towards the bluer part of the spectrum. The model continuum can vary unevenly with wavelength in the infra-red and hence there is a possibility of the spectrum being improperly normalised even when we divide it by the global highest value of the continuum during injection of the model. This means that the continuum will not be set exactly to 1. To serve as a visual aid, we further divide each row by its median while leaving out any masked data to account for these residual normalisation issues. The results of this study do not change if this step is skipped. If the SVD/PCA based detrending were ideal, there should be no effect on the retrieved injected signal at this stage. In reality, although we do recover the injected signal, we also see that it has been affected by the detrending process, as depicted in Panels (c) and (d) of Figure \ref{fig:steps} with their cross-sections at close to mid-transit shown in Panels (f) and (g). It is evident that the spectrum in Panel (g), while having the lines at the same positions as in Panel (f), is attenuated in depth and shows an alteration of the line depth and line wing shapes. Previous results in literature have also claimed that the detrending/processing algorithm distorts any actual signal present in the data. \citealt{brogi2019retrieving} show the effect of SVD based detrending on an injected emission model extending over the entire spectral sequence rather than just a few spectra as we do here for transit, while \citealt{meech2022applications} showcase dampening of a model signal through both SYSREM and airmass linear regression based detrending methods. This study, while complimenting the results above about the presence of detrending induced distortion of the signal, expands the context of distortion to a signal present only in a few in-transit phases rather than the full sequence of spectra. These effects are necessary to account for, since we shall be using the log-likelihood approach in our analysis (see Section \ref{25}) which is very sensitive to all these parameters \citep{brogi2019retrieving}.
\\
\\
There is an inherent assumption that the noiseless data matrix would have flux from almost all other non-moving sources except the moving exoplanet. However, the exact variations captured depends on the number of SVD components used to create the noiseless matrix. In addition to the effects due to processing on the model itself in Panel (d) of Figure \ref{fig:steps}, we also observe slight \emph{excesses} causing the presence of artefacts in the out-of-transit phases. A cross-section of the artefacts at a chosen out-of-transit phase is shown in Panel (h). If the SVD/PCA based detrending had been perfect, these out-of-transit values would all be 1.0. However, we instead observe some excesses that are greater than 1.0. The maximum variation above 1.0 for such excesses is about 3.3 times lesser than the maximum depth of the line cores of the injected exoplanet signal at the mid-transit phase, as shown in Panel (f). \citet{brogi2019retrieving} also had artefacts being produced as a result of SVD based detrending, but it was for the case of an injected emission model spectrum extending over the entire spectral sequence and the artefacts were produced around the reprocessed model to also extend over the entire sequence. This study now shows the case for a transmission model injection within a few spectra only, but the artefacts still seem to cover the entire spectral sequence rather than just the reprocessed model in transit. While the exact cause of such artefacts is yet to be determined, these structures should also be present in the analysis of actual data. We test this hypothesis and its possible use via injection tests in Section \ref{31}.

\subsection{The log-likelihood approach}\label{25}
Now that both the data and the model have been processed with the processing bias also accounted for (in the case of the model), we need to calculate a Bayesian estimator which will help us do parameter estimation as well. For this, we use the \emph{Likelihood} ($L$) function as defined in \citet{brogi2019retrieving} to calculate the log-likelihood value as:
\begin{equation}
    \log(L) = -\frac{N}{2}\log[s_{f}^{2} + 2R(s) + s_{g}^{2}].
\end{equation}
Here, $N$ is the number of unmasked pixels, per row as we compare a row per order of the original processed data (\textbf{A$_\mathrm{n,d}$}) to the corresponding row per order in the processed model (\textbf{A$_\mathrm{n,m}$}). The variance of the data ($s_{f}^{2}$), variance of the model ($s_{g}^{2}$) and the cross-covariance ($R(s)$) are calculated as:
\begin{eqnarray}
    s_{f}^{2} & = & \frac{1}{N}\sum_{n}f^{2}(n), \nonumber \\
    s_{g}^{2} & = & \frac{1}{N}\sum_{n}g^{2}(n), \nonumber \\
    R(s)      & = & \frac{1}{N}\sum_{n}f(n)g(n-s). 
\end{eqnarray}
Here, $f(n)$ and $g(n)$ are the mean subtracted values of a row from each order of the processed data matrix (\textbf{A$_\mathrm{\textbf{N},d}$}, see Section \ref{23}) and the corresponding row from each order from the processed model matrix (\textbf{A$_\mathrm{\textbf{N},m}$}, see Section \ref{24}) respectively. In previous literature using CCFs as the analytical tool, the cross-correlation coefficient was calculated as $C(s) = R(s)/\sqrt{s_{f}^{2}s_{g}^{2}}$ \citep{brogi2019retrieving} and provides the basis behind the name of this approach as CCF-to-$\log(L)$. $s$ denotes a shift in wavelength, which is obtained by Doppler-shifting the model for each tested value of radial velocity. Thus, the value of $s$ changes depending on the particular exoplanet orbital solution used for injection (see Equation 5 in Section \ref{24} for parameters used here) before processing the model. For our analysis, $v_\mathrm{rest}$ (which is simply $v_\mathrm{P}-v_\mathrm{sys}$) is calculated on a grid between $-20$ and 20 km\,s$^{-1}$ in steps equally spaced by 1\ km\,s$^{-1}$; $K_\mathrm{P}$ is computed between 70 and 150 km\,s$^{-1}$ in intervals of 1 km\,s$^{-1}$. Processing all the models corresponding to this grid, calculating the $\log(L)$ for each row for the orders considered (in both the processed data and model cuboids) and then summing across rows followed by summing across orders to calculate a single $\log(L)$ value for each model would then leave us with a matrix of $\log(L)$ values corresponding to each orbital solution (each $v_\mathrm{rest}$-$K_\mathrm{P}$ pair). The highest value of $\log(L)$ should ideally fall at the expected exoplanet orbital solution in the case of a detection. For GJ 3470\,b, the expected $K_\mathrm{P}$ is:
\begin{equation}
    K_\mathrm{P} = \frac{2\pi a}{P_\mathrm{P}\sqrt{1-e^{2}}} \sin(i),
\end{equation}
where $a$ is the semi-major axis, $P_\mathrm{P}$ is the orbital period and $i$ is the orbital inclination. Using value of $P_\mathrm{P}$ as $3.336649^{+0.000084}_{-0.000081}$ days from Section \ref{introduction}, value of $a$ as $12.92^{+0.72}_{-0.62}$ $R_{\star}$ (for $R_{\star}$ = $0.48^{+0.04}_{-0.04}$ $R_{\odot}$) \citep{kosiarek2019bright}, the value of $i$ as 89.13$^{\circ +0.26^{\circ}}_{-0.34^{\circ}}$ \citep{awiphan2016transit} and assuming a circular solution, the predicted value of $K_\mathrm{P}$ is $94.1^{+8.6}_{-8.6}$ km\,s$^{-1}$ (using multivariate functional approach for error propagation and using only the largest errors for each variable to propagate). If we were to include the effects of $e$ by assuming a value of $0.114^{+0.052}_{-0.051}$ \citep{kosiarek2019bright}, $K_\mathrm{P}$ would be $94.7^{+8.6}_{-8.6}$ km\,s$^{-1}$. This would also induce an additional radial velocity shift of about -1.8 km\,s$^{-1}$ at transit mid-point compared to a circular orbital solution (for the case of the first night of observation). Since both the $K_\mathrm{P}$ values are very close and the induced radial velocity shift at transit mid-point is small, we just assume a circular solution for our analysis (as mentioned in Section \ref{24} as well). However, we make sure to have a large enough velocity grid in $K_\mathrm{P}$-$v_\mathrm{rest}$ space for the \emph{Model Selection} procedure in Section \ref{33} to also account for these velocity differences.
\\
\\
From the $\log(L)$ matrix, we can calculate confidence interval maps via a likelihood ratio test. Assuming that the maximum $\log(L)$ within the matrix is $\log(L_\mathrm{max})$, the \emph{Likelihood ratio statistic} ($\lambda$) is:
\begin{equation}
    \lambda = 2[\log(L_\mathrm{max}) - \log(L)].
\end{equation}
Wilks' Theorem \citep{wilks1938large} then states that with a large enough sampling, $\lambda$ would approach a $\chi^{2}$ distribution with two degrees of freedom in our case ($v_\mathrm{rest}$ and $K_\mathrm{P}$). The Survival Function of this statistic (p-value matrices) corresponding to a $\chi^{2}$ distribution is then calculated\footnote{\tt{scipy.stats.chi2.sf($\lambda$, 2)}}. The confidence interval values are then calculated by finding the corresponding Inverse Survival Functions corresponding to a Normal distribution\footnote{\tt{scipy.stats.norm.isf(p-value/2)}}. Under this criterion, the maximum $\log(L)$ value would have confidence interval equal to zero ($\lambda=0$) by construction, and the standard error on the measured velocities is given by the 1-$\sigma$ contour. Generally, the constraints on $v_\mathrm{rest}$ are tighter compared to $K_\mathrm{P}$.
While the likelihood ratio test does not allow us to quote a ``detection significance'', a signal confidently detected would appear as a series of tight, concentric contours around a certain velocity pair, and all the rest of the parameter space disfavoured by more than 4-5$\sigma$, the exact threshold depending on how conservative one chooses to be regarding detecting a species 
(see Section \ref{31} for an example).
However, sometimes spurious features mimicking a detection 
can come up at positions corresponding to telluric contamination and aliasing. In such cases, it is best to compare how well the detection itself stands out in comparison to the other spots.
\\
\\
This same approach can also be extended if the likelihood function is computed on a grid of models generated using differing chemical abundances and cloud top pressures and is used for our Model Selection maps in Section \ref{33}. In this case, models (i.e. combination of parameters) corresponding to the $n$-$\sigma$ confidence interval are rejected at the same confidence level with respect to the model with the highest likelihood.

\subsection{Forward Models}\label{26}

\begin{figure}
    \centering
    \includegraphics[width = \columnwidth]{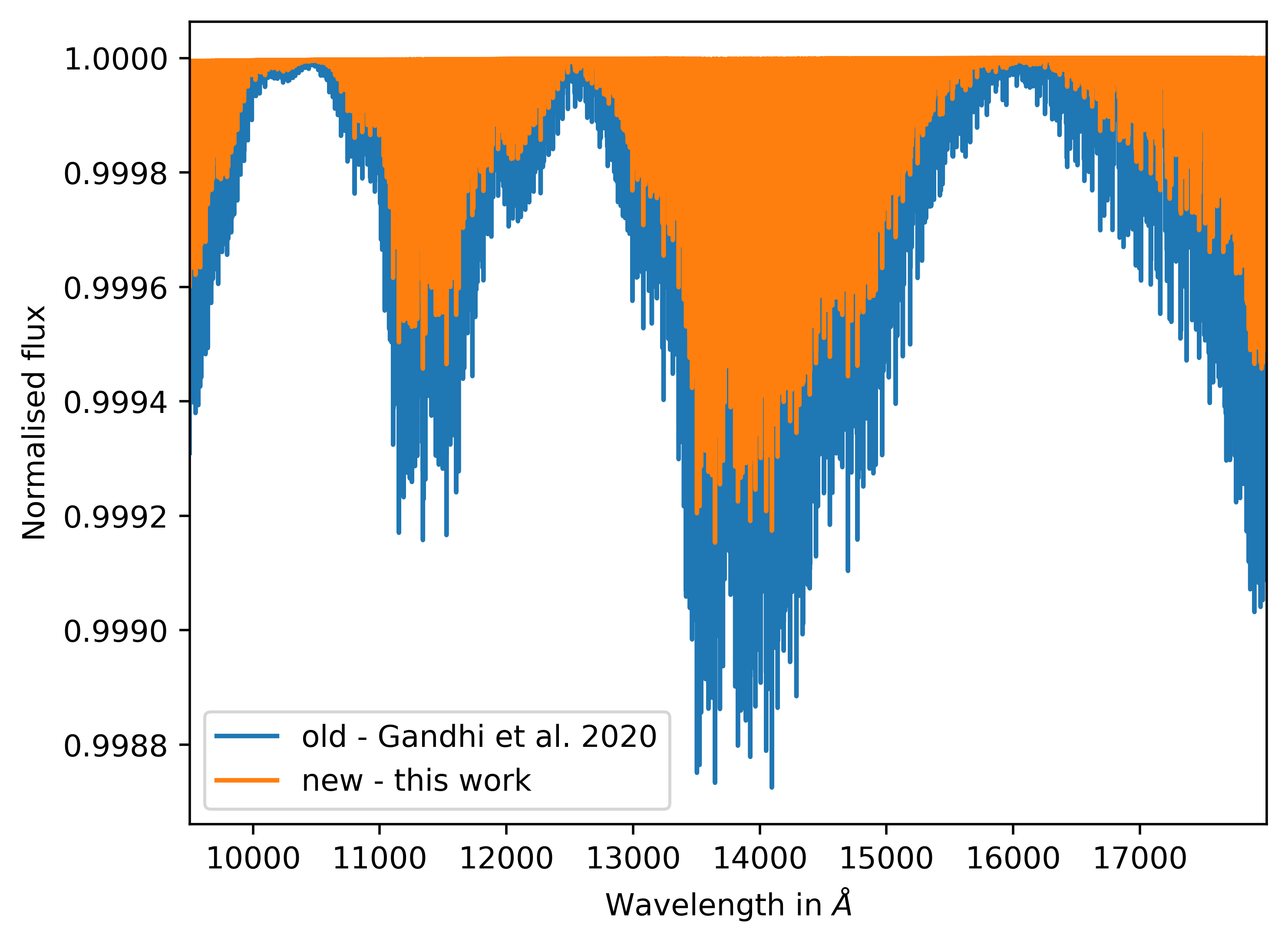}
    \caption{Comparison between the normalised old best fit model to low resolution HST WFC3+Spitzer observations in \citet{benneke2019sub} calculated in \citet{gandhi2020seeing} (in blue) and the corresponding revised model calculated in this work (in orange). For details regarding the models please see Section \ref{26}. The new model has a noticeable reduction in line core strength (on median about 2.2 times less compared to the old model) which has consequences to their detectibility (see Section \ref{31}).}
    \label{fig:modelcompare}
\end{figure}

Models for the atmosphere of GJ 3470\,b were computed using the GENESIS framework adapted to transmission spectroscopy \citep{gandhi2017genesis, pinhas2018retrieval, pinhas2019h2o} with a grid of pressure between 10$^1$ and 10$^{-8}$~bar, assuming hydrostatic equilibrium and abundances of molecular species that are constant with altitude. An optically-thick layer of clouds is simulated by setting the optical depth to infinity below a characteristic cloud-top pressure $P_\mathrm{C}$. With this assumption, we are not testing the possible effects of Mie scattering (i.e. a wavelength-dependent aerosol opacity) hinted by \citet{benneke2019sub}, because the wavelength range of CARMENES does not extend enough to overlap their claimed opacity drop long-ward of 2 micron.
We explored three different scenarios for our models:
\begin{itemize}
    \item We initially ran injection tests and searched for a real exoplanet signal by using the best-fit model of \citet{gandhi2020seeing}, that is $\log_{10}$(H$_2$O) = $-3.0$ (for Volumetric Mixing Ratio (VMR) abundance, to be assumed as the default notation for writing abundance values henceforth) and $\log_{10}(P_\mathrm{C}) = -2.3$. This model is shown in blue in Figure~\ref{fig:modelcompare} (``old model'').
    \item We then constructed a grid of models in ($\log_{10}$(H$_2$O), $\log_{10}P_\mathrm{C}$), both equally spaced by 0.5 dex between $-5$ and $0$, and with a pressure-temperature profile ($P$-$T$) broadly matching \citet{benneke2019sub}. The latter is parametrised with 3 pressure-temperature pairs, namely ($P_0, T_0$) = (10 bar, 1000 K), ($P_1, T_1$) = (0.1 bar, 650 K), and ($P_2, T_2$) = (10$^{-3}$ bar, 450 K). The atmosphere is assumed isothermal above (below) the maximum (minimum) pressures above, while in-between a constant lapse rate is assumed. The line list used for H$_2$O is POKAZATEL from ExoMol \citep{tennyson2016exomol, polyansky2018exomol, gandhi2020molecular}. Figure \ref{fig:modelcompare} (orange lines) shows the closest match to the \citet{gandhi2020seeing} spectrum mentioned above, with parameters ($\log_{10}$(H$_2$O) = $-3.0$, $\log_{10}(P_c) = -2.5$). The revised $P$-$T$ profile results in a noticeable reduction in the line strength (on median about 2.2 times or 46.2\% less), compared to the previous work where the atmosphere was isothermal at the planet's equilibrium temperature. Such change in line strength has consequences on the estimated detectability of signals via injection tests (Section \ref{31}).
    \item Including only water and a cloud deck might not be representative of the atmosphere of the planet if other minor species are present and can alter the continuum level. Therefore, we also generated a grid akin to the previous one, but with the VMR of CH$_4$ set to 10$^{-5}$ and that of NH$_3$ to 10$^{-4.5}$, qualitatively matching the upper limits reported in \citet{benneke2019sub}. The line lists used for CH$_4$ and NH$_3$ are the ones produced in \citet{hargreaves2020accurate} (for HITEMP) and \citet{coles2019exomol} (for ExoMol) respectively \citep{gandhi2020molecular}. These additional species make a noticeable difference only in the case of very-low cloud deck and very low water abundance (See Section \ref{42}). 
\end{itemize}

\section{Results}\label{results}
\subsection{Injection Testing}\label{31}
\begin{figure*}
    \centering
    \includegraphics[width = \columnwidth]{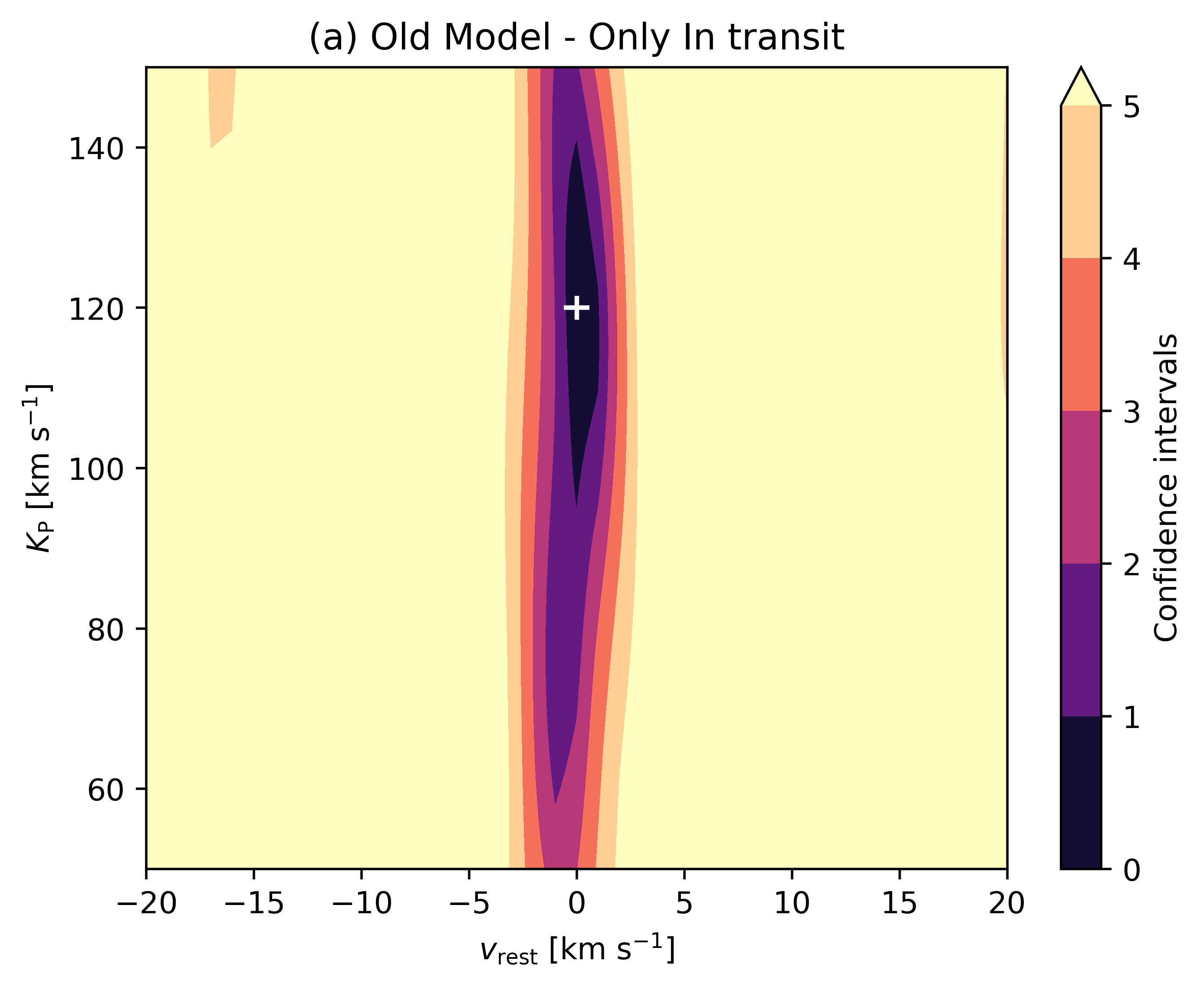}
    \includegraphics[width = \columnwidth]{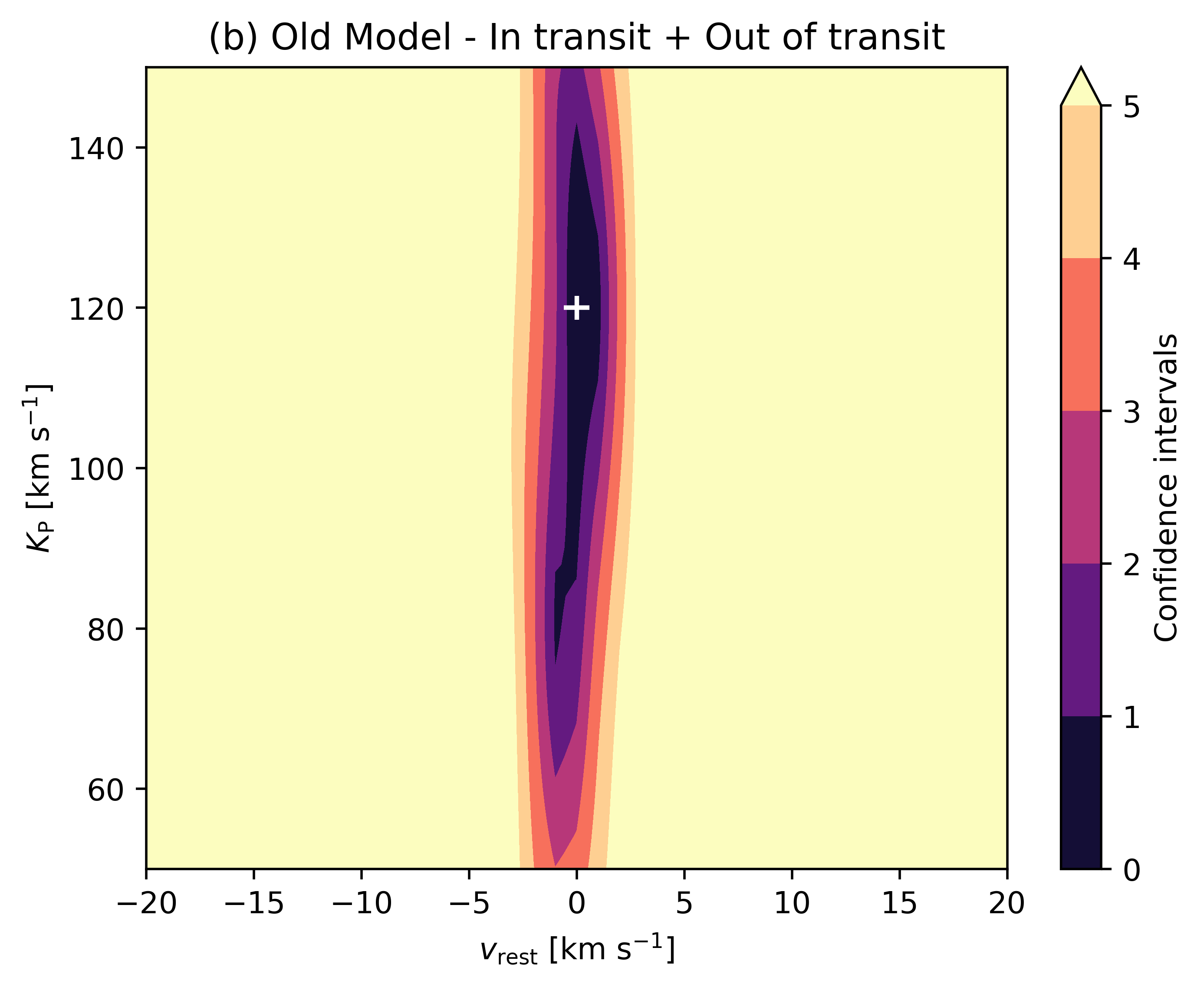}
    \includegraphics[width = \columnwidth]{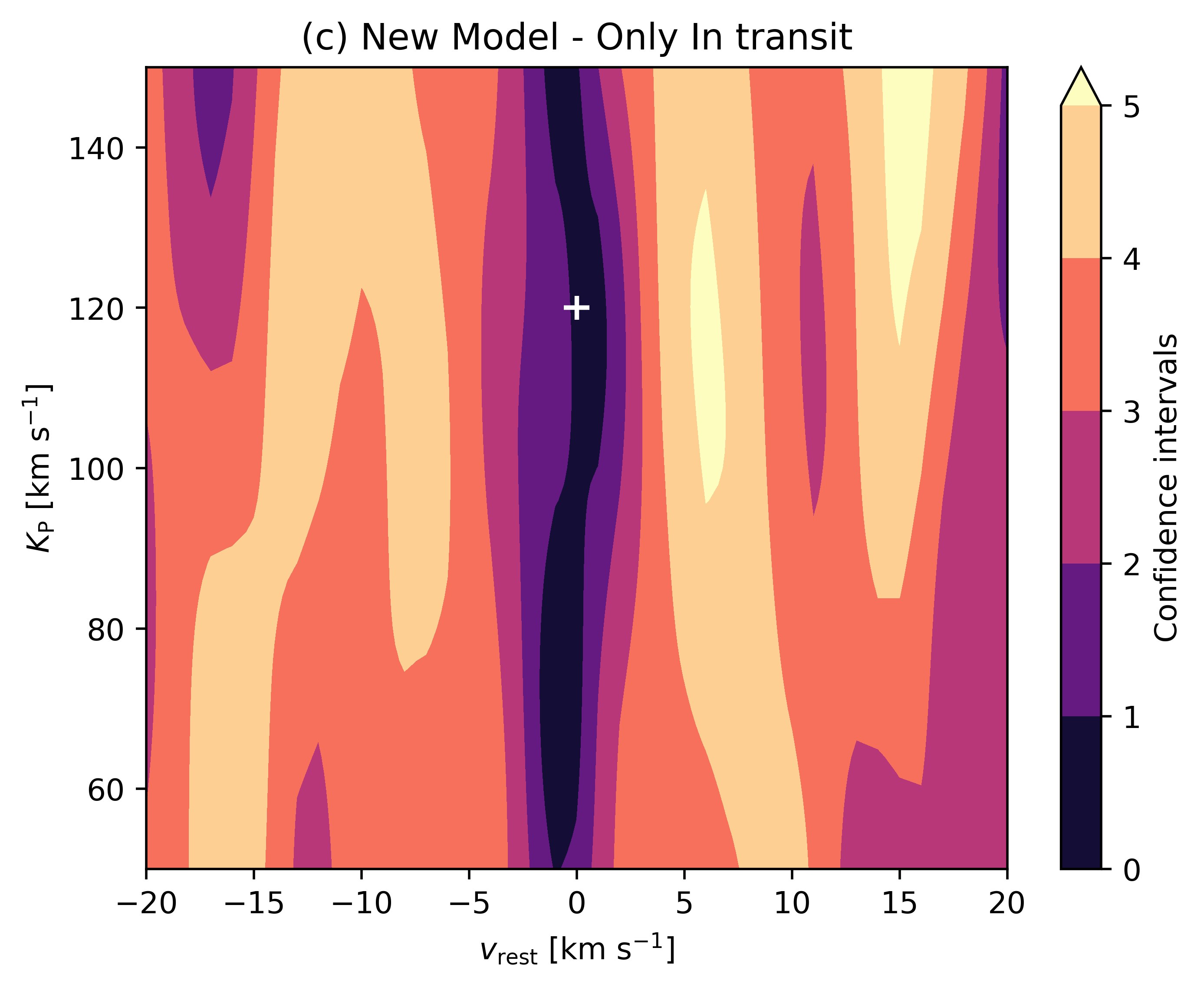}
    \includegraphics[width = \columnwidth]{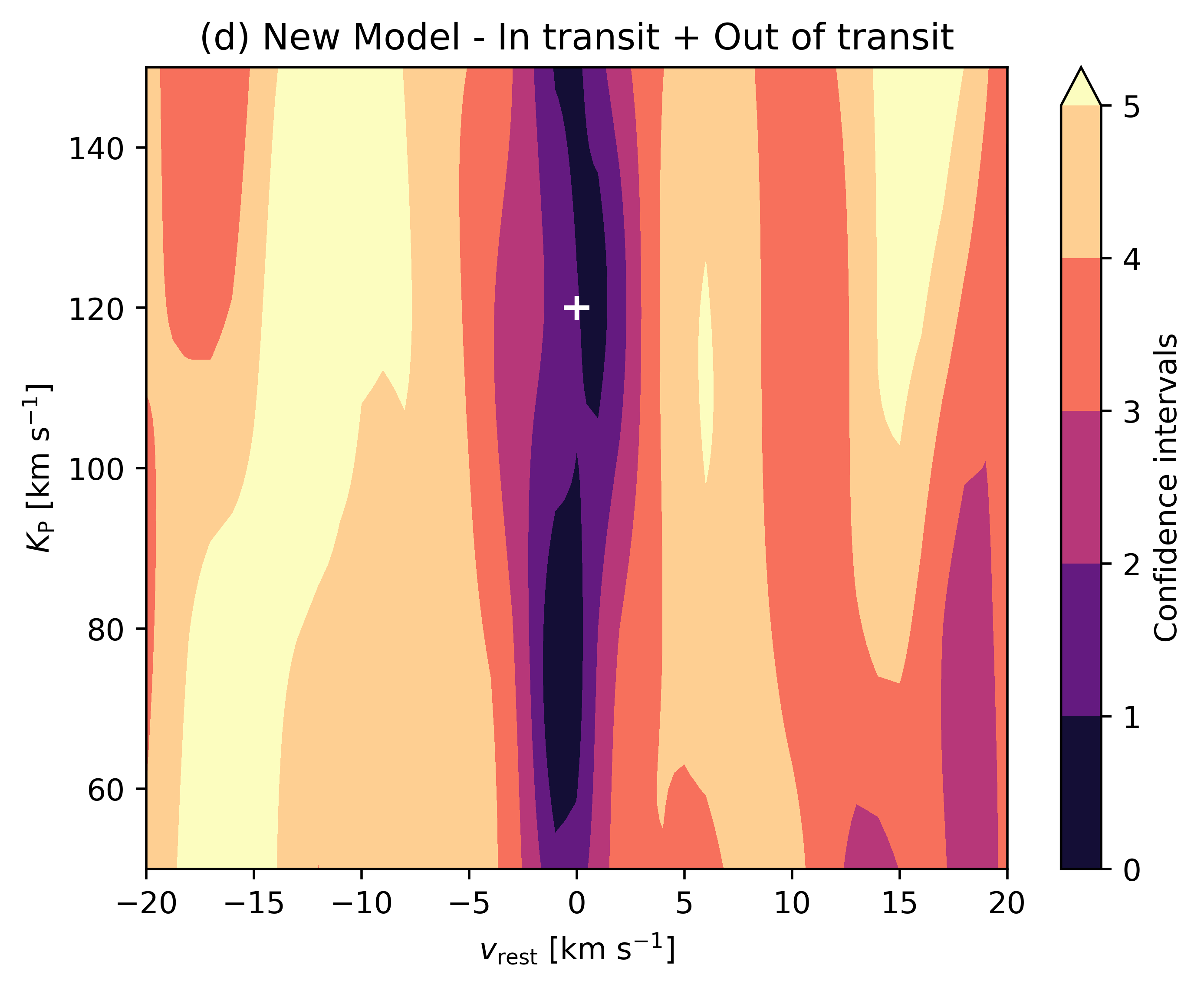}
    \caption{Recovery of a H$_2$O signal injected at the nominal level, on a $v_\mathrm{rest}$ - $K_\mathrm{P}$ grid, using two nights of observations. The white plus marker shows the location of the injected model for each night and the grid of $v_\mathrm{rest}$ values have been accordingly shifted to be centred around $v_\mathrm{sys}$ of this injected signal. \textbf{(a)} For the model used in \citet{gandhi2020seeing} which we denote as ``Old Model" and by cross-correlating only the in-transit portions of the spectral sequences. \textbf{(b)} Same as (a) but now by cross-correlating the entire spectral sequences, including out-of-transit spectra. \textbf{(c)} Same as (a) but with the model we generate in this work (denoted as ``New Model"). The difference in line strength manifests with no visible detection in this case compared to (a). \textbf{(d)} Same as (b) but with the New Model, which now shows some signature of the injected signal but doesn't stand out much in comparison to a similar feature in its neighbourhood. Overall including the out-of-transit data with the artefacts seems to slightly increase the level of significance of the detection for a nominal injected signal.}
    \label{fig:injecttestdetect}
\end{figure*}
To test if our pipeline works as intended and to predict the sensitivity to atmospheric scenarios, we inject a nominal exoplanet signal in each night at $v_\mathrm{sys}$ = -10 km\,s$^{-1}$ and $K_\mathrm{P}$ = 120 km\,s$^{-1}$. Since, $v_\mathrm{sys}$ for this system is 26.09 km\,s$^{-1}$ \citep{brown2018gaia}, the injected signal is 36.09 km\,s$^{-1}$ blue-shifted from the expected systemic velocity of any possible exoplanet signal and hence should result in minimal overlap if there's already an actual signal present in our data. We then see if we can detect this injected signal by combining both nights. As discussed in Section \ref{26}, there are two best fit models we want to test out - the model from \citet{gandhi2020seeing} and the revised (shallower) model generated for this work. We use 6 components for the SVD/PCA based detrending step throughout our analysis since we find that a nominal injected signal equivalent to the model used in \citet{gandhi2020seeing} (``old model", which is the only model successfully detected by our pipeline) is detected  at this point. We also find that Night 1 from our observations is marked by heavy telluric line saturation in the orders 8-10, 15 and 18-21 and some CO lines in orders 24 and 25. We do not use these for our analysis. Night 2 was of better quality and hence we only remove orders 9, 15, 18-19 and 24-25 from our analysis.
\\
\\
We show the confidence interval plots for each injected model in Figure \ref{fig:injecttestdetect}. Panels (a) and (b) are for the case of nominal injection of the \citet{gandhi2020seeing} model and Panels (c) and (d) are for the case of nominal injection of the revised model. Panels (a) and (c) show the case when we use just the in-transit spectra to calculate the log-likelihoods and Panels (b) and (d) are for the case of using the entire processed matrix (including the out-of-transit spectra as well as the in-transit spectra) to calculate the log-likelihoods. We remind that the latter strategy is worth testing due to the appearance of spurious, correlated structures out-of-transit, due to the application of SVD/PCA based detrending. It is possible to detect the model of \citet{gandhi2020seeing} unambiguously ((at $>5\sigma$) from the injection using two nights if we use the entire matrix (Panel (b)). It is also possible to do the same by just using the in-transit spectra (Panel (a)). On the other hand, the revised model calculated in this work presents only a weak detection (at $\sim3\sigma$) when we use the entire matrix (Panel (d)), compared to a non-detection in Panel (c) where we use only the in-transit spectra. This is expected because as we saw in Figure \ref{fig:modelcompare}, the more accurate new best fit model has much shallower line depths compared to the previous model.
\\
\\
We discussed the effect of model reprocessing by our pipeline in the out-of-transit data in Panel (h) of Figure \ref{fig:steps} where we saw that there were slight deviations from 1.0 in the residuals. From Panel (b) and (d) in Figure \ref{fig:injecttestdetect}, we also see that including the effect of such artefacts (in addition to the actual signal) while calculating likelihood values seems to yield slightly stronger detections. This might seem counter-intuitive because adding the out-of-transit spectra should not add more information about the signal itself and hence should dilute the magnitude of detection. However, if our model reprocessing accurately mimics the effects of the data analysis on the observations, we expect these out-of-transit residuals to appear also in the presence of a real exoplanet signal in-transit. Furthermore, it seems likely that the shape and slope of the artefacts themselves  is also unique for each orbital solution in the grid. Hence, using the whole spectral sequence rather than just the in-transit spectra might allow us to pinpoint the solution more precisely because we now cross-correlate the artefacts in the reprocessed model with the similar level of artefacts produced in the out-of-transit spectra for the observed data cube as well after processing. We will test whether this approach introduces biases on the overall retrieval process in Section \ref{33}. We then discuss the results from both this section (Section \ref{31}) and Section \ref{33} together in Section \ref{41} as part of the Discussion.

\subsection{On real observations of GJ 3470\,b}\label{32}
\begin{figure*}
    \centering
    \includegraphics[width = \columnwidth]{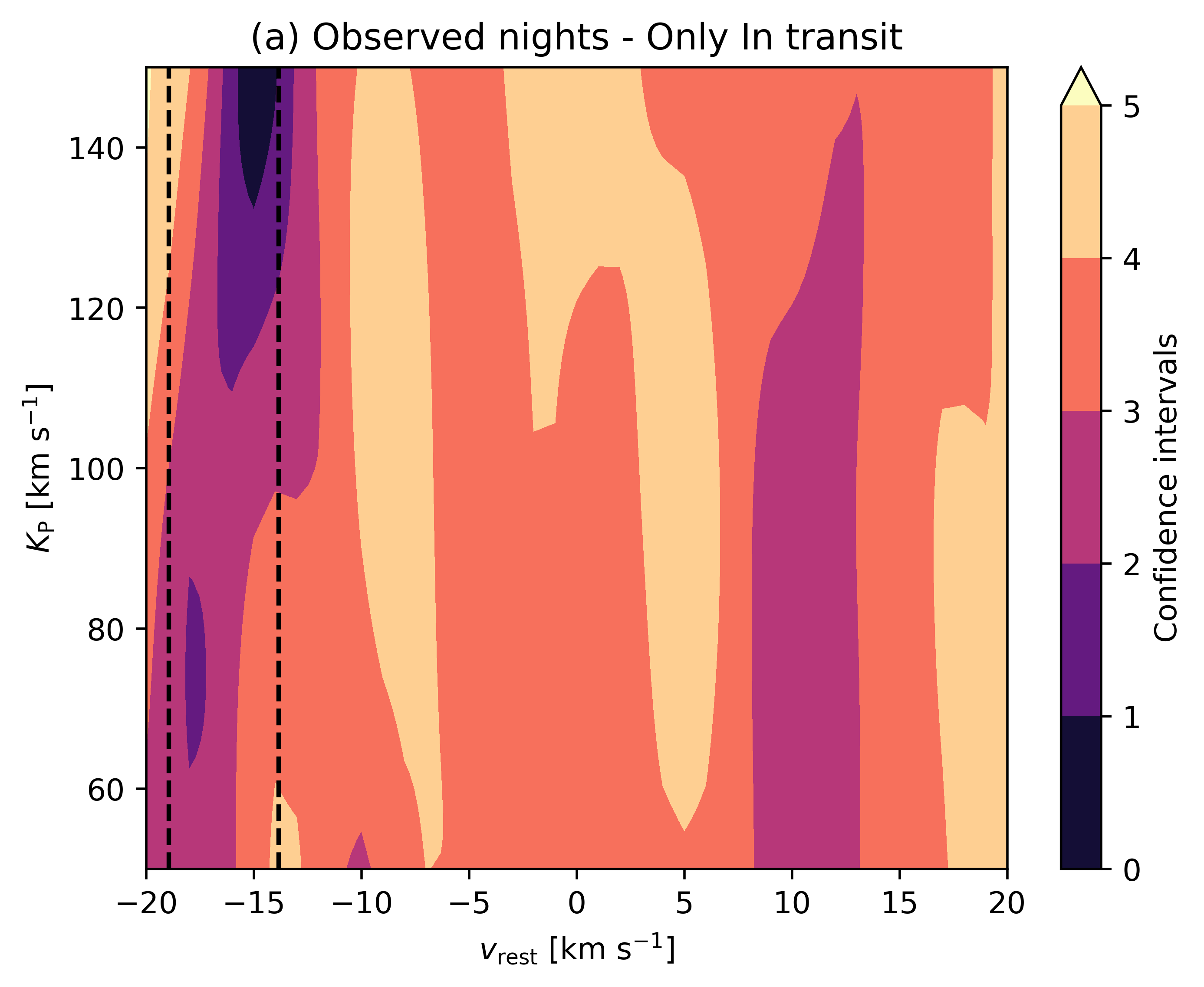}
    \includegraphics[width = \columnwidth]{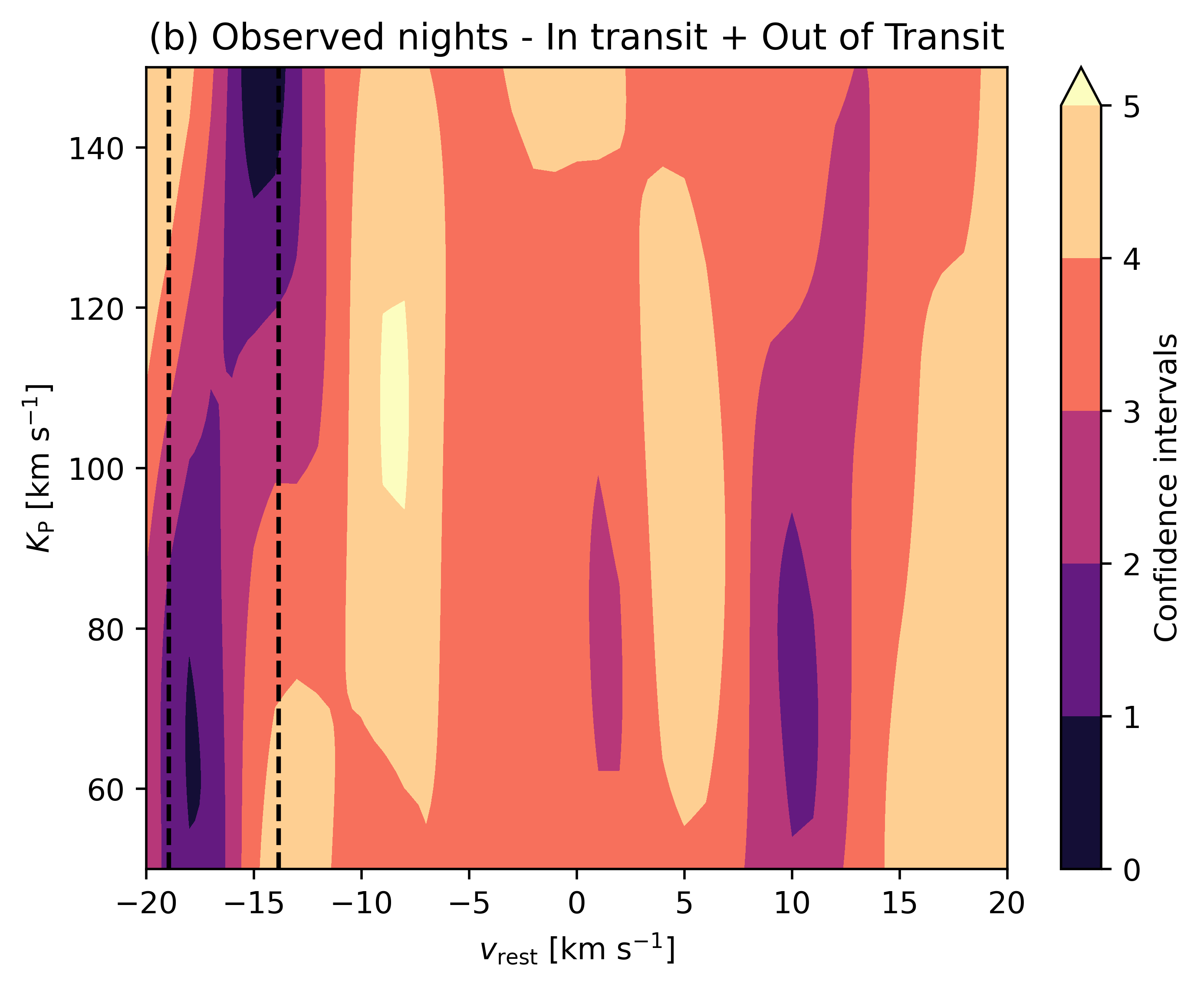}
    \caption{Same as Figure \ref{fig:injecttestdetect} but now with only the two observed nights (with no nominal signal injection). Both cases do not show any detection of a H$_2$O signature matching the revised best fit model to \citet{benneke2019sub} calculated in this work (New Model). The black dashed lines represent the velocities at which we expect contamination due to telluric residuals to appear (between v$_\mathrm{rest}$ of -13 and -19 km\,s$^{-1}$).}
    \label{fig:rawtestdetect}
\end{figure*}

We repeat the same process as above with 6 SVD/PCA components and the same orders removed from each night's analysis, but now with just the two nights of observed data with no model injections. The confidence interval plots are shown in Figure \ref{fig:rawtestdetect}. Panel (a) shows the case of cross-correlating just the in-transit spectra for the processed data and reprocessed model, and Panel (b) for the case of cross-correlation of the entire matrices. Neither case shows a detection.
Both plots however show some common features that fall roughly between rest frame velocities of -13 to -19 km\,s$^{-1}$. Any features due to telluric contamination for each night are expected to be visible at $v_\mathrm{rest}$ $=$ average($v_\mathrm{bary}$) - $v_\mathrm{sys}$. The features falling between -13 and -19 km\,s$^{-1}$ pointed above hence very well coincide with the calculated velocities at which telluric contamination is expected to happen for both nights (-13.85 km\,s$^{-1}$ for night 1 and -18.94 km\,s$^{-1}$ for night 2), as marked by the black, dashed lines. The presence of these uncorrected traces of telluric contamination shows that even with a very aggressive pipeline to remove telluric lines and mask strong residuals, residual tellurics still persist below the level of the noise, and at the level comparable to the exoplanet signal itself.
\\
\\
A non-detection in these plots indicates that either the signal is too weak to be detected, or the model tested does not have the correct line strength - especially because the likelihood value calculation is very sensitive to a global scaling. Works like \citet{guilluy2022gaps} have introduced a \emph{scaling factor} to account for such cases where they attenuate or inflate their model by a factor before cross-correlating and calculating the likelihoods values. Coupling this with an MCMC algorithm would then allow us to get the value of the scaling factor at which the likelihood calculation at the expected signal position is maximised. 
\\
\\
Doing a similar calculation to obtain a scaling factor in this work is not appropriate because the models used here include the effects of a cloud deck pressure level as well unlike the ones used in \citet{guilluy2022gaps}. Higher cloud decks for any fixed abundance would mean attenuation of line depths. This means that the cloud deck pressure layer would be degenerate with the effect of a model with no clouds but with a decreasing scaling factor. However, this motivates us to perform a model selection on a grid of models with varying H$_{2}$O abundance and cloud deck pressures to constrain any possible signal - which we denote in the following sections as \emph{Model Selection}. This kind of retrieval was also used in \citet{lafarga2023hot} where they used it to constrain H$_{2}$O abundance using ESPRESSO data in the optical for the sub-Neptune WASP 166\,b. In this work, we extend this framework to analyse results from infra-red observations as well.

\subsection{Constraints on the cloud deck pressure level and water vapour content}\label{33}
\begin{figure*}
    \centering
    \includegraphics[width = \columnwidth]{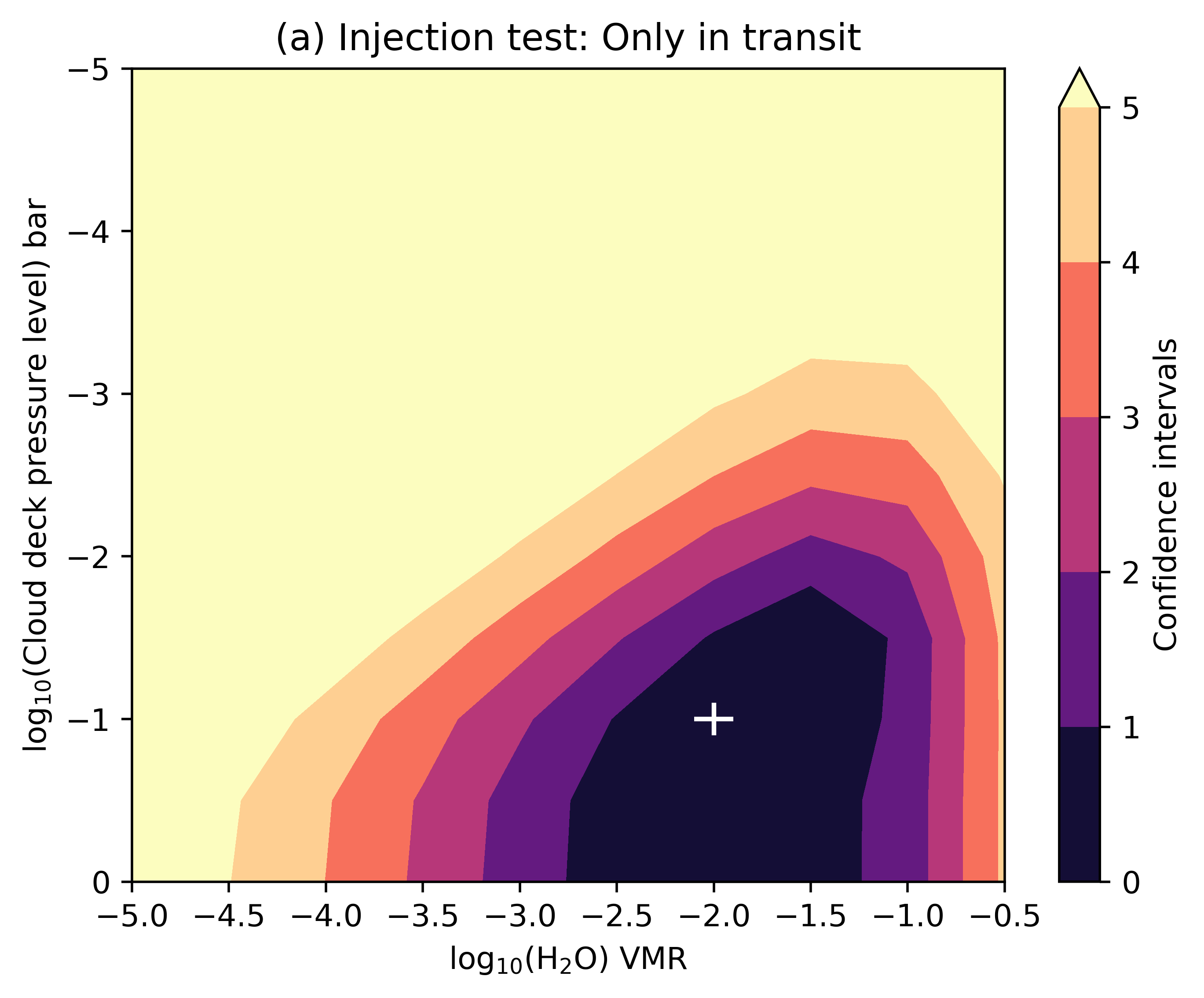}
    \includegraphics[width = \columnwidth]{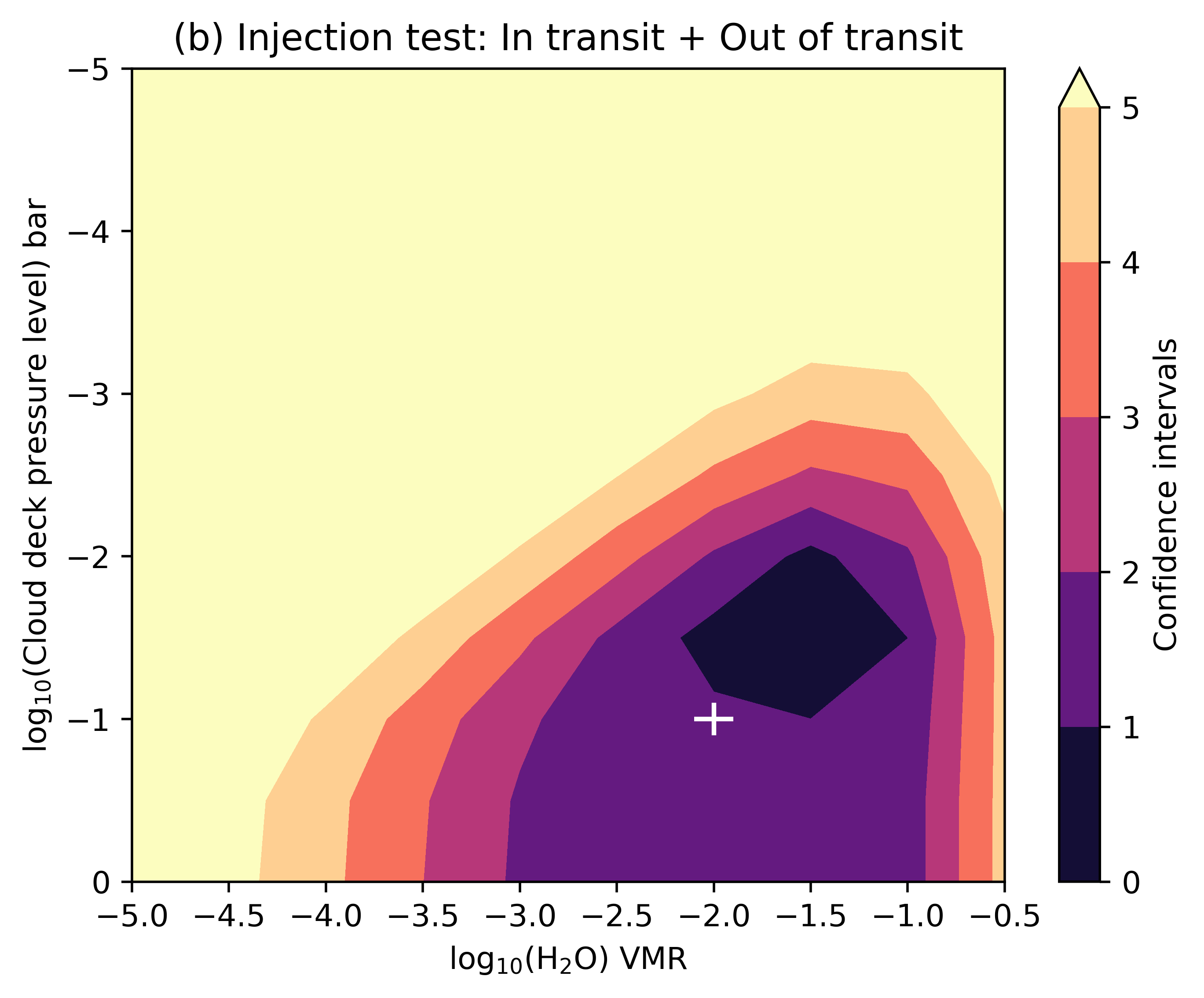}
    \includegraphics[width = \columnwidth]{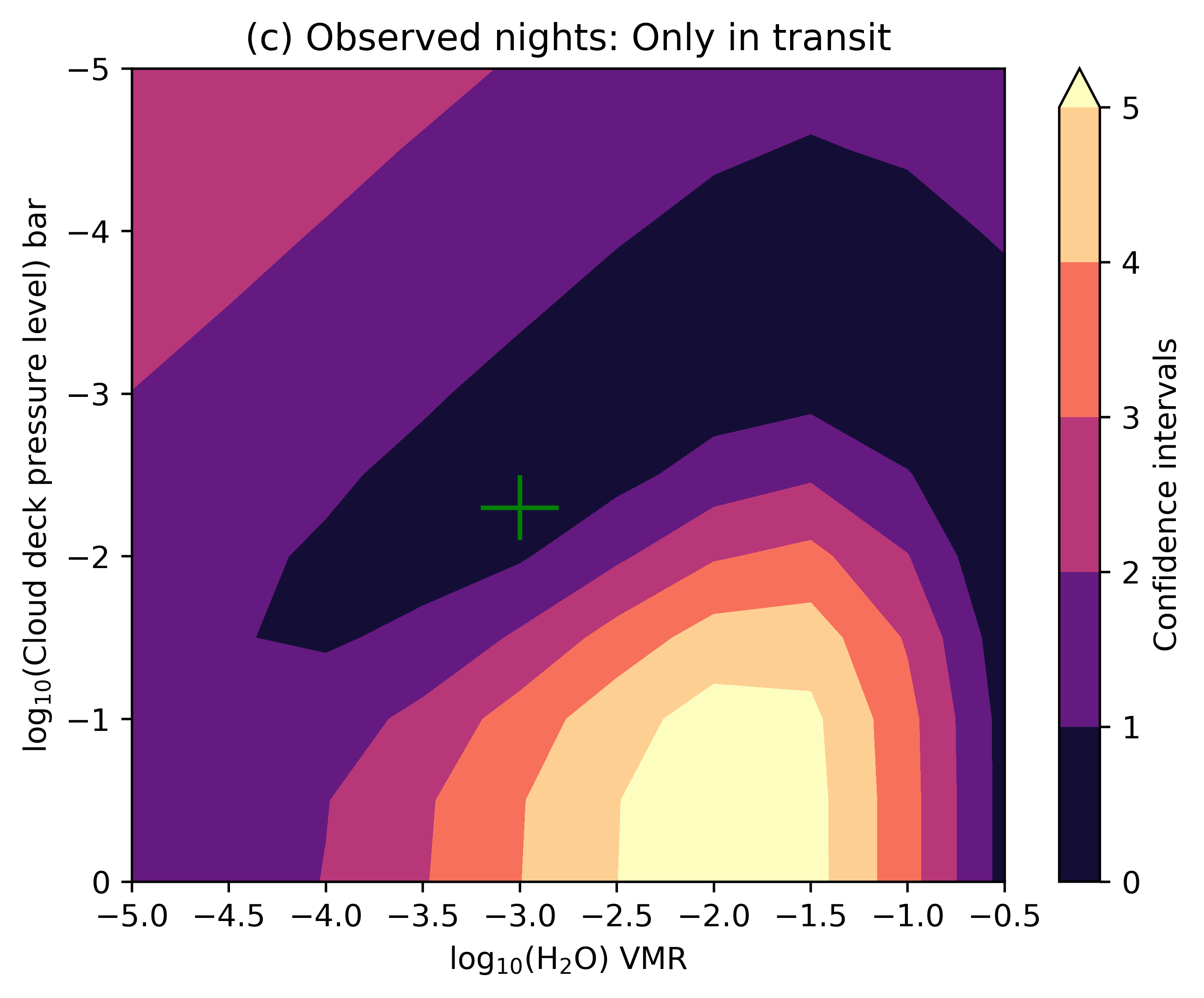}
    \includegraphics[width = \columnwidth]{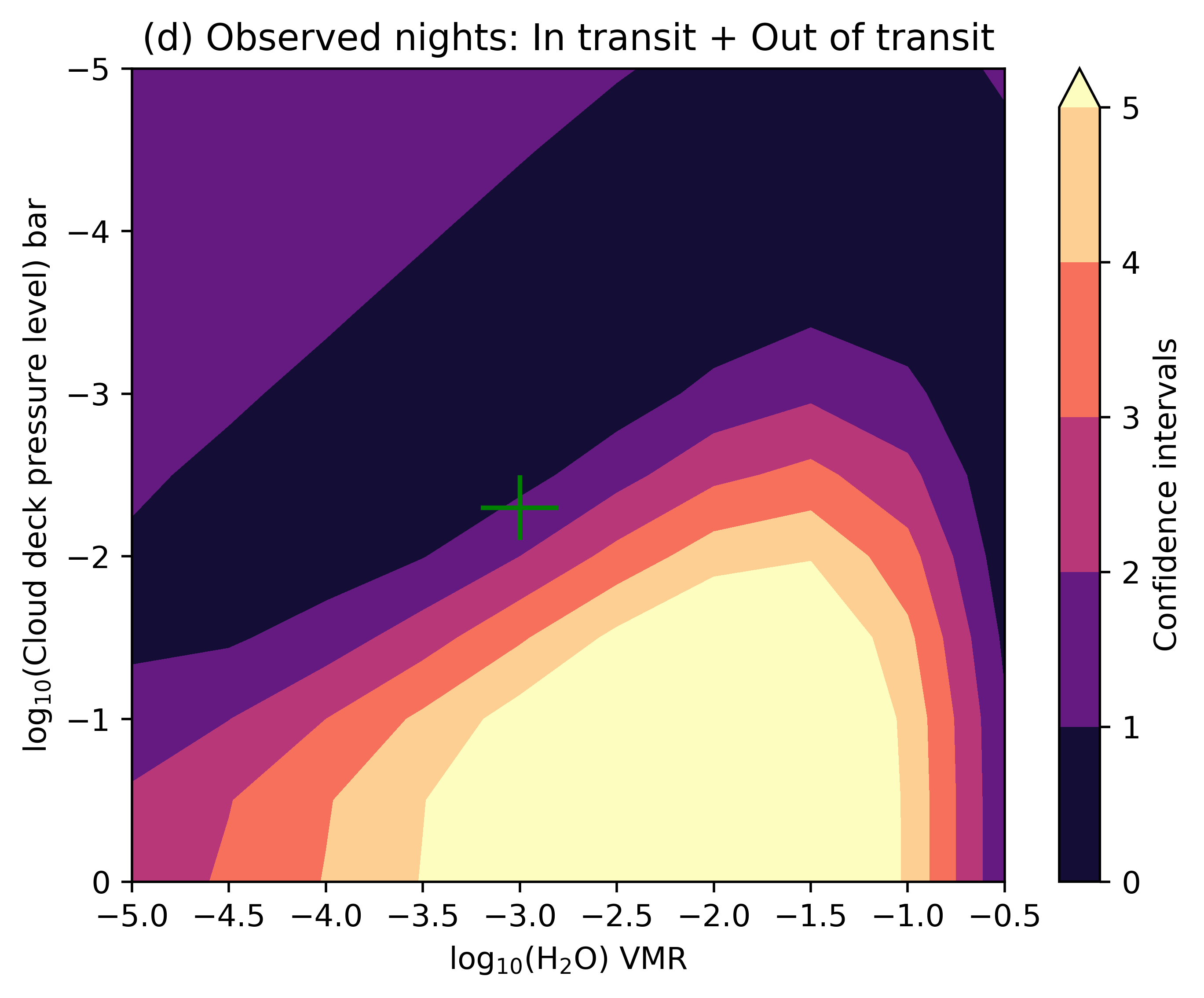}
    \caption{Model Selection on a grid of H$_2$O abundance (increasing when we move right in the grid) versus cloud deck pressure level (decreasing when moving upwards in the grid) using two observed nights with CARMENES. \textbf{(a)} and \textbf{(b)} present the case of selection after a nominal injection of a signal at log(H$_2$O) $=$ -2.0 and log(P) = -1.0 in both nights (white plus marker, also please see Section \ref{33} for details). (a) uses only the data within the in-transit phases for calculating Likelihoods while (b) uses the entire dataset including the out-of-transit phases as well. Both cases successfully retrieve the injected signal 
    but (b) has its 1$\sigma$ \textbf{contour} region slightly up and to the right of the actual values of the injected signal. This could show that using the entire dataset can lead to some biases for this process. \textbf{(c)} and \textbf{(d)} are obtained by only using the observed nights with no model injection and represent possible constraints on a real but marginal at best H$_2$O signal in the data. The strip within the 1$\sigma$ contours represent the degenerate models that will showcase similar line core strengths. Both (c) and (d) share the 1$\sigma$ contour regions and hence the results are consistent. However, (d) excludes more cloud free scenarios but also cannot exclude a case of a flat line for the upper left region of the plots. The green plus marker shows the approximate location for the best-fit abundance and cloud deck constraints obtained from HST WFC3+Spitzer.}
    \label{fig:injecttestselection}
\end{figure*}

Motivated by the framework established in \citet{lafarga2023hot}, in this section we perform a \emph{Model Selection} on a grid of models with varying H$_2$O abundances and cloud deck pressure levels. The grid of models used to constrain the atmosphere of GJ 3470\,b is described in Section \ref{26}. For each of the models, we repeat the procedure done for the two nights of actual observations in Section \ref{32} (with the same number of SVD/PCA components) but now limit our $v_\mathrm{rest}$ values to be between [-5, 5] km\,s$^{-1}$ with spacing of 2 km\,s$^{-1}$, and $K_\mathrm{P}$ values to be between [70, 130] km\,s$^{-1}$ with spacing of 4 km\,s$^{-1}$. This is done to keep computational times reasonably manageable and to focus only in the velocity space where the exoplanet signal should arise, thus avoiding the strong telluric contaminants highlighted in Section \ref{31}. Furthermore, the calculated theoretical $K_\mathrm{P}$ of GJ 3470\,b would fall roughly in the middle of the range, which is big enough to account for shifts due to the exoplanet mass being slightly inaccurate in literature or eccentricity playing a role. Ultra-hot and hot Jupiters have had yielded detections highly shifted from $v_\mathrm{rest}$ $=$ 0 due to the presence of strong winds in the atmospheric terminators. However, we do not anticipate shifts greater than 1 km\,s$^{-1}$ due to winds in exoplanets like GJ 3470\,b \citep{landgren2023shallow,innes2022atmospheric}. After calculating the log-likelihoods on the reduced $v_\mathrm{rest}$ and $K_\mathrm{P}$ matrix defined above, for each model, we store the highest value of the likelihood values in the grid. This value then populates each point of the grid seen in both panels in Figure \ref{fig:injecttestselection}. The same likelihood-to-confidence intervals approach outlined in Section \ref{25} is used to calculate the confidence interval contours for our model selection. The likelihood grid are calculated for each night first and then added before we do build the combined confidence interval plots that we see in Figure \ref{fig:injecttestselection}.
\\
\\
The easiest models to detect would be the ones with the deepest line cores. For the model grid we have assumed, this would correspond to cloud-free or low cloud deck (i.e. clouds are much deeper in the atmosphere) models (a high cloud deck would result in highly attenuated line cores) with not too high $\log_{10}$(H$_{2}$O) abundance (too high would result in high mean molecular weight highly compact atmospheres, which would again result in attenuated line cores). Hence, we next see if our pipeline can successfully retrieve such a model. We inject a nominal model with $\log_{10}$(H$_2$O) $=$ -2.0 (super-solar, since solar value is -3.0) and $\log_{10}(P)$ = -1.0 at $v_\mathrm{sys}$ = -10 km\,s$^{-1}$ and $K_\mathrm{P}$ = 120 km\,s$^{-1}$ (same as the case for injection testing in Section \ref{31}) in both nights of observations. The velocity grid we use for calculating likelihoods is thus accordingly shifted to be centred around this injected signal.
\\
\\
Panels (a) and (b) in figure \ref{fig:injecttestselection} show the combined results of our \emph{Model Selection} test on the injected nights. It is easy to see that if we cross-correlate only the spectra in transit (Panel (a)), we can recover our model successfully within the 1$\sigma$ contour. If we cross-correlate the entire night of both the processed data and the re-processed model (Panel (b)), the constraints of all contours become slightly tighter, especially the 1$\sigma$ contour which now becomes highly localised around $\log_{10}$(H$_2$O) $=$ -1.5 and $\log_{10}(P)$ = -2.0. However, the 1$\sigma$ contour now lies slightly away from the model parameters we actually injected the nights with (which still lies within the 2$\sigma$ contour). The 1$\sigma$ localised model parameters should have degenerate line core strengths with our injected model, hence it is definitely interesting to understand why the selection process decided to converge on that model specifically in the case of cross-correlating both the in-transit and out-of-transit data. One possible reason could be that the calculation with entire matrices is slightly biased because of the presence of artefacts, or presence of random noise can drag the maximum likelihood to a different position other than the injected signal in some cases. However, this still doesn't drastically change the nature of the results.
\\
\\
Panels (c) and (d) in Figure \ref{fig:injecttestselection} show the combined results of a \emph{Model Selection} test on the two observed nights by cross-correlating just the spectra in transit and cross-correlating entire matrices respectively. From the outset, it is easy to see that both analyses decisively exclude the region which would be the easiest to detect (see discussion on the retrieval of an injected model above) at >5$\sigma$. The 1$\sigma$ contours of both are also consistent. Panel (c) - with cross-correlation for the in-transit spectra only - favours a family of degenerate models, which spans from very high abundance ($\log_{10}$(H$_2$O) = -0.5) and cloud-free models (highly compressed heavy atmospheres), to super-solar abundances (-2.5$<\log_{10}$(H$_2$O)$\leq$-1) at high cloud deck models (-2.5$>\log_{10}(P)\geq$-4.5), and then to slightly super-solar and sub-solar  abundances (-4.5$<\log_{10}$(H$_2$O)$\leq$-2.5) with moderately high cloud deck models (-1.5$>\log_{10}(P)\geq$-2.5). Interestingly, the region corresponding to the weakest possible spectra (low water abundance and high clouds), which is the closest scenario to a flat spectrum and thus a non-detection, is also excluded at a significance between 2$\sigma$ and 3$\sigma$. This suggests that - despite not being able to produce a solid detection in ($v_\mathrm{rest}$-$K_\mathrm{P}$) as shown in Figure \ref{fig:rawtestdetect}, the data does contain some level of correlated signal, enough to marginally exclude the flattest scenarios.
\\
\\
In Panel (d) where both in-transit and out-of-transit spectra are cross-correlated, the 1$\sigma$ contour extends to even more sub-solar abundances at moderate cloud deck models and to super-solar abundances at very high cloud deck models in addition to what Panel (c) encompasses. The level at which Panel (d) rejects cloud free models across all H$_2$O abundances is higher than in Panel (c), but on the flip side it is no longer able to reject very weak models (with low abundances and very high cloud decks) at the same level since they now fall within the 2$\sigma$ contour.
\\
\\
As with the case of the injected tests in Panels (a) and (b) of Figure \ref{fig:injecttestselection}, the results of Panels (c) and (d) for just the observed nights are consistent. As a whole, both analyses together show that GJ 3470b is highly unlikely to be an exoplanet with cloud-free atmosphere, but they can only marginally reject the case of the transmission spectrum being completely flat, which would be characterised by very high cloud deck levels for almost all H$_2$O abundances in the model grid we constructed, or just indicate that there is no signal. Hence, both our analyses suggest that the measurement of H$_2$O from both nights of observations is only marginal at best. It is however notable that the results from low resolution observations of GJ 3470b from HST WFC3 + Spitzer in \citet{benneke2019sub} (solar H$_2$O abundance and cloud deck $\log_{10}P_\mathrm{C}$ = -2.3) falls within the 1$\sigma$ contour in Panel (c) and in the 2$\sigma$ contour in Panel (d), and is hence still consistent with our results.

\section{Discussion}\label{discussion}
\subsection{Using out-of-transit spectra to help pinpoint the exoplanet signal}\label{41}
As discussed in Section \ref{22}, the decision to process the entire night of observation rather than just the data in-transit was because of the limit imposed on the SVD/PCA based detrending approach due to the number of spectra available. However, this leads to production of artefacts in the out-of-transit phases around the actual injected model signal during model reprocessing. As these artefacts should also be present in real data, we tested whether they could be used to precisely and accurately derive atmospheric properties. In order to do so, we used injection tests to study the difference in results when we use just the in-transit data for log-likelihood calculation versus using the entire matrix (in-transit + out-of-transit data).
\\
\\
As seen in the injection tests in Section \ref{31} (or Figure \ref{fig:injecttestdetect}), the addition of the out-of-transit phases produces confidence intervals that are marginally tighter. 
While this result might appear contradictory, we note again that in this case the ``artefacts'' are in fact produced as a result of the presence of a signal in the in-transit phases. Hence, each orbital solution should also have a correspondingly unique artefact signature and adding the out-of-transit phases as well would lead to additional information to support the presence of the signal as well. This is due to an increase in cross-correlation between the signal artefacts with the model artefacts, in addition to the cross-correlation of the actual signal with the reprocessed model.
\\
\\
An increase in detection significance however manifests slightly differently when it comes to Model Selection plots (Figure \ref{fig:injecttestselection}). From Figure \ref{fig:injecttestselection} (a) and (b), we see that while the constraints on the contours for the retrieval of an injected model are stronger when the full matrix is cross-correlated, it also led to the 1$\sigma$ contours falling slightly away from the actual injected signal. The fact that we cross-correlate artefacts in the case of full matrix cross-correlation in (b) makes it possible that there is cross-correlation of artefacts from sources which are not the actual signal but anything else that can mimic it. Possible sources could include contamination due to stellar lines from the M-dwarf source, telluric contamination (both at the level of the signal and hence difficult to remove by our SVD/PCA based detrending approach), and could also just be due to random noise since we are at the low S/N limit with these observations. This also makes the case for finding an optimum number of SVD/PCA components to use to reduce this possible contamination. While this approach doesn't substantially change the nature of retrieved results in our case (which shows the efficiency of our SVD/PCA based detrending approach), it still has the potential to mislead and must hence be noted in future studies.
\\
\\
For the case of Model Selection on only the real data, with no injection (Panels (c) and (d) of Figure \ref{fig:injecttestselection}), while both methods lead to overlapping 1$\sigma$ contours the full matrix case (Panel (d)) leads to a larger parameter space enclosed within the same contours. However, the excluded parameter spaces at a 3$\sigma$ limit also increase to cover almost all cloud-free models in the grid. Hence, while the result of the retrieval again doesn't change in just the observed nights, a full matrix case is able to exclude a larger parameter space and hence has the same effect as in the injection model retrieval case. However, since any possible signal in the data is weak, it is also no longer able to constrain the limit of cloud deck level and is hence not able to reject the case of a flat line model at the upper left corner of the grid at >3$\sigma$. Hence, on a weak or no signal, the case of utilising the entire matrix now constrains the result to a hard lower limit on cloud deck pressure level by rejecting almost all cloud-free model cases compared to the case of only using the in-transit data.
\\
\\
All in all, these preliminary tests on the utilisation of the full spectral sequence rather than just the in-transit portion of the observations seem to prove that the results are not particularly biased and at least similarly precise. Further work is required to assess whether this result can be extended to any transit observation or is rather just a particular outcome of this data set.
\\
\\
While we chose to operate in the time domain to align with the majority of the literature to date, we also tested whether these artefacts persist when repeating the detrending procedure in the wavelength domain, with the differences between the two approaches highlighted in Section \ref{22}. We do not observe any similar out-of-transit excesses in this latter case. However, the quality of the telluric correction for the same number of SVD/PCA components appear inferior, and telluric residuals persist after the correlation stage. While it is certainly worth noting the difference between time and wavelength domain, a thorough comparison between the two approaches is beyond the scope of this work. We present the analogue to Figure \ref{fig:steps} but in the wavelength domain in Figure \ref{fig:steps_wavdom} in the appendix.

\subsection{Comparison with previous results}\label{42}
In Section \ref{32} and Figure \ref{fig:rawtestdetect}, we show that we are unable to detect a water vapour signature matching low resolution observations from HST+WFC3 and Spitzer as done in \citet{benneke2019sub}, who found instead a >5$\sigma$ detection of H$_2$O. Their detection and the corresponding retrieved atmosphere forms the basis of the model grid we computed and tested in this paper. 
\\
\\
Our non-detection with just two CARMENES transits is somewhat expected, as \citet{gandhi2020seeing} predicted that 4 nights of observations of GJ 3470\,b using either of CARMENES, GIANO or SPIRou would be needed to detect the best fit model they generated. In this work we see that the model they used is too optimistic based on the \citet{benneke2019sub} results, which would then push the number of nights of observations needed to an even higher number. Even though the approach presented in this paper would be sufficient to detect the model used in \citet{gandhi2020seeing} in just 2 nights, we still see that it is not enough to detect the revised and more realistic but weaker model (see Figure \ref{fig:modelcompare}). 
\\
\\
Interestingly, even without any significant detection, we are still able to place some constraints on the H$_2$O abundance and cloud deck pressure level, with both nights showing consistent behaviour and overall rejecting almost all cloud free models across the tested grid. The 1$\sigma$ contours also overlap with the results from the low resolution observations of \citet{benneke2019sub}; however the area within the 1$\sigma$ contours from our analysis is very broad and extends across a range of high abundance low cloud deck pressure models (very high mean molecular weight atmospheres) to low abundance and moderate cloud deck models. Unfortunately, the upper limit of the cloud deck pressure level is still unconstrained. All taken together, while the present analysis doesn't necessarily suggest the presence of H$_2$O in the atmosphere of GJ 3470\,b, the constraints it can lay on the retrieved parameters seem promising. This motivates future analysis with more high resolution data sets.
\\
\begin{figure*}
    \centering
    \includegraphics[width = \columnwidth]{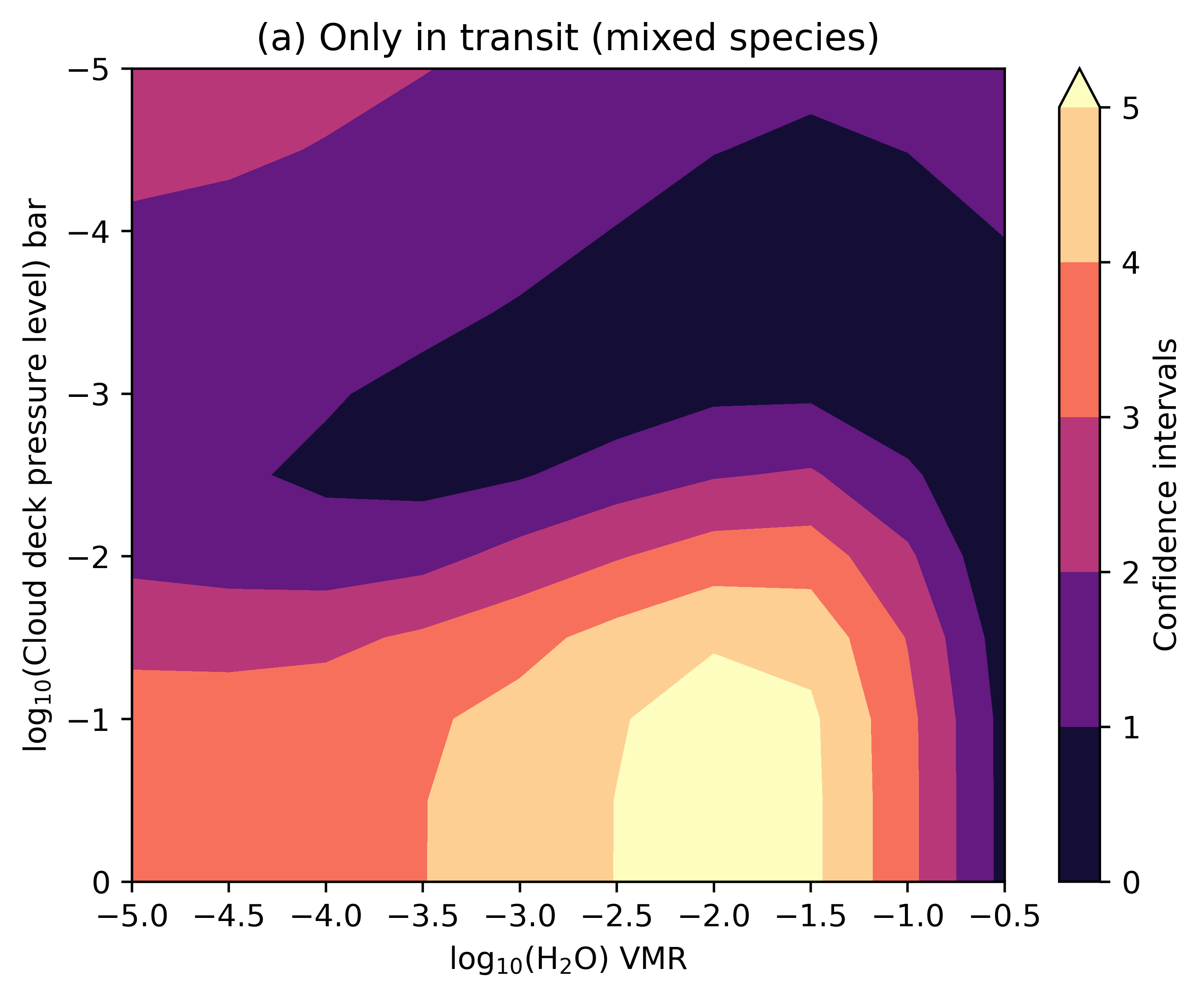}
    \includegraphics[width = \columnwidth]{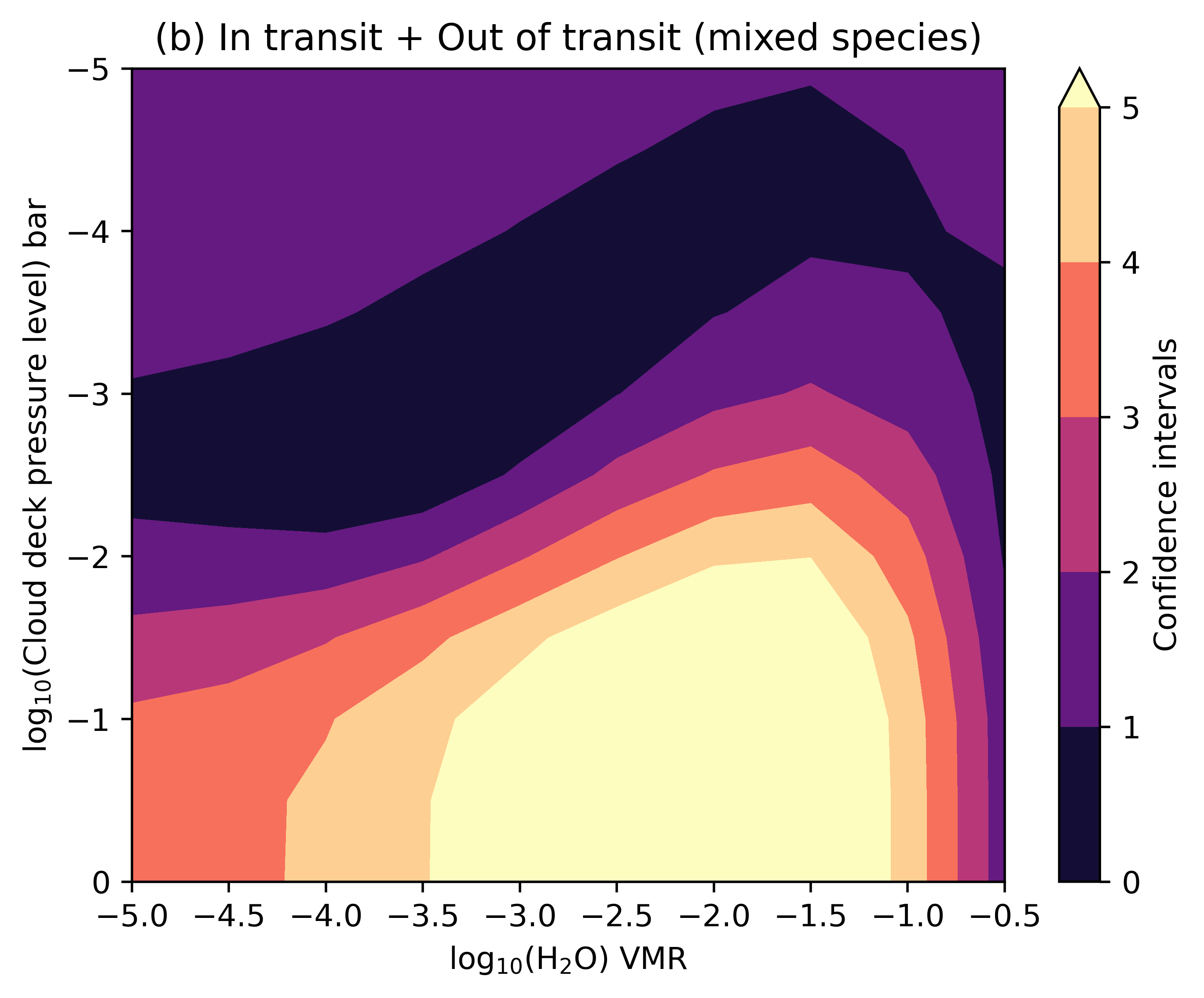}
    \caption{Same as Panels (c) and (d) from Figure \ref{fig:injecttestselection} but now with CH$_4$ and NH$_3$ included at their upper limit values obtained from \citet{benneke2019sub}. The overall result within the 1$\sigma$ contours doesn't change much from their counterparts in Panels (a) and (b) from Figure \ref{fig:modelcompare} but more models in the lower left corner (low H$_2$O abundance and cloud free/ deep cloud deck models) are rejected to a greater extent. Hence addition of more molecules increases the possibility of the atmosphere to be necessarily cloudy compared to the results obtained with just H$_2$O.}
    \label{fig:mixedsp}
\end{figure*}
\\
An additional complication that could be present for smaller and cooler objects like GJ 3470\,b is that searching for individual molecules using HRCCS in the infra-red in exoplanetary atmospheres could prove more difficult than in the case of larger and hotter objects. We restricted our work to the search of only H$_2$O because its dominant presence was suggested by low resolution results. The same results in \citet{benneke2019sub} also suggested a depletion of CH$_4$ and provided upper limits to log(VMR-abundances) of both CH$_4$ and NH$_3$ as $\approx$ -5 and -4.5 respectively. CH$_4$ is expected to be present in the atmospheres of such objects and hence its depletion in low resolution results was notable. To see if the inclusion of some additional molecules has any effect on the model selection plots, we repeated the analysis in Section \ref{33} but now instead use a grid of models which also include CH$_4$ and NH$_3$ at the levels suggested above in addition to H$_2$O (See Section \ref{26} for details). The revised plots are shown in Figure \ref{fig:mixedsp}. Compared to their counterparts in Panels (c) and (d) in Figure \ref{fig:modelcompare}, both Panels (a) and (b) show an increased rejection of models in the lower left hand corner i.e. low H$_2$O abundances and cloud-free or deep cloud deck cases. These areas also coincide with the region in which the effect of the additional molecules on the model spectra starts becoming significant due to reduced contribution of the dominant H$_2$O feature. Rejection of these mixed species models suggests that their presence would then only be possible with higher cloud decks. Nevertheless, the overall result denoted by the regions inside 1$\sigma$ contours do not change much compared to the case with only H$_2$O included and hence the discussion in the preceding paragraph still remains valid. The main takeaway from this result should be that addition of other species can serve to break the degeneracy as well. Observations using other instruments with wider wavelength coverage like GIANO and/or SPIRou would make it possible to include molecules like CO and CO$_2$ as well for a similar analysis, although it would also require the implementation of a full MCMC (rather than a grid exploration) due to the increased number of parameters.

\subsection{Further work}\label{43}
High resolution spectroscopic analysis involves cross-correlating a forest of observed lines with model template lines to compare line positions, and with the additional log-likelihood approach, also with the lines' depth and shape. Hence, it is imperative that any model templates used for comparison with observations be as accurate as possible. Otherwise, the analysis will simply fail to report a signal even in the presence of an actual signal in the data. \citet{guilluy2022gaps} had already indicated that in smaller and cooler exoplanetary candidates, disequilibrium chemistry might start having a very prominent role compared to hot and ultra-hot Jupiter cases. In this work, for generating the best fit template models as well as the model grid used for model selection, a free chemistry assumption was used. While allowing the abundances to adjust freely allows for more flexibility, the abundances are assumed to be constant with altitude, which might simply be inaccurate if disequilbrium chemistry is present. However, looking at changes due to assumption of disequilibrium chemistry (instead of free) is out of scope of this work, but is something that could be looked at in the future. One challenge in such an assumption would be the computational resources needed. Hence, observations of GJ 3470\,b taken using the JWST (GTO Program 1185, PI: Greene\footnote{From https://www.stsci.edu/~nnikolov/TrExoLiSTS/JWST/trexolists.html}) could already establish a foundation to build upon, in the same way we used the low resolution results from HST WFC3+Spitzer observations to start our analysis from.

\section{Conclusion}\label{conclusion}
In this study we analyse two nights of publicly available data sets obtained using the CARMENES spectrograph mounted at CAHA with the aim of detecting signatures of H$_2$O in the atmosphere of the sub-Neptune GJ 3470\,b:
\begin{itemize}
    \item We modified an existing SVD/PCA based detrending approach to include the effects of the detrending process on any exoplanet planet signal present in those data sets through Model Reprocessing (see Figure \ref{fig:methodology} for a schematic overview of the entire process). 
    \item We found that the number of exposures in each observed night presented a limit to the SVD/PCA based detrending process which resulted in us processing the entire night of observation rather than just the data predicted to have been taken within the exoplanet transit phases. However, this also caused the production of correlated artefacts in the out-of-transit spectra during the model reprocessing procedure (Figure \ref{fig:steps}), which led us to investigate whether these out-of-transit artefacts contain meaningful and unbiased signatures of the exoplanet atmosphere. 
    \item Using the entire sequence of phases rather than limiting our cross-correlation to just the in-transit phases produced a marginally stronger detection in injection tests. However, such tests also revealed the potential for a small bias (between 1$\sigma$ and 2$\sigma$) in the derived abundances and cloud top-pressure for strong signals.
    \item Using the two nights of CARMENES observations, we were unable to detect a signal by cross correlating with a spectrum matching previous low-res observations of \citet{benneke2019sub}. Extending the templates to a grid of models with varying H$_2$O abundances and cloud deck pressure does not lead to a detection either (Figure \ref{fig:rawtestdetect}).
    \item 
    In spite of the non detection, 
    a likelihood ratio test performed on the above grid of models 
    favoured a family of degenerate models with similar line core strengths (see Figure \ref{fig:injecttestselection}).
    \item The results retrieved from low resolution observations in \citet{benneke2019sub} (solar H$_2$O abundance and cloud deck $\log_{10}P_\mathrm{C}$ = -2.3) fall within the 1$\sigma$ contour of this work and are hence still consistent with our results from high resolution observations. However, our analysis is unable to either strongly reject completely flat spectrum or place strong constraints on the parameters, which indicates that the presence of any actual H$_2$O signal is only marginal at best.
    \item To see if inclusion of more species can affect this result, we included CH$_4$ and NH$_3$ at the upper abundance limits calculated in \citet{benneke2019sub}. We don't see much change in the Model Selection (see Figure \ref{fig:mixedsp}) result except for exclusion of even more cloud-free models at low H$_2$O abundances where both molecules would start competing with H$_2$O for dominance in the spectra (which is obviously not the case even from low resolution observations).
\end{itemize}
In this work we only analysed two nights of CARMENES observations as they were already publicly accessible. During this work, there have been observations of GJ 3470\,b taken using GIANO and SPIRou, but they are still in proprietary mode. The Bayesian approach developed in this work makes it feasible to combine analysis from observations made using different instruments with diverse resolutions (including observations from JWST as well). Hence, future work could look into combining multiple observations across a variety of ground and space based spectrographs to understand the challenges associated with such an approach and to see if that enhances or challenges the results of this work.

\section*{Acknowledgements}

SD is supported by a Chancellors' International Scholarship from the University of Warwick. SG is grateful to Leiden Observatory at Leiden University for the award of the Oort Fellowship. ML acknowledges funding from the UKRI (grants MR/S035214/1 and EP/X027562/1). A portion of this research was carried out at the Jet Propulsion Laboratory, California Institute of Technology, under a contract with the National Aeronautics and Space Administration (80NM0018D0004).

\section*{Data Availability}

The data used in this work can be downloaded freely from the Calar Alto archive\footnote{http://caha.sdc.cab.inta-csic.es/calto/jsp/searchform.jsp}. The Python scripts used for data reduction and analysis can be shared by the authors upon reasonable request.




\bibliographystyle{mnras}
\bibliography{bibliography} 



\appendix

\section{Detrending in the wavelength domain} \label{appendix}

\begin{figure*}
    \centering
    \includegraphics[width = 2\columnwidth]{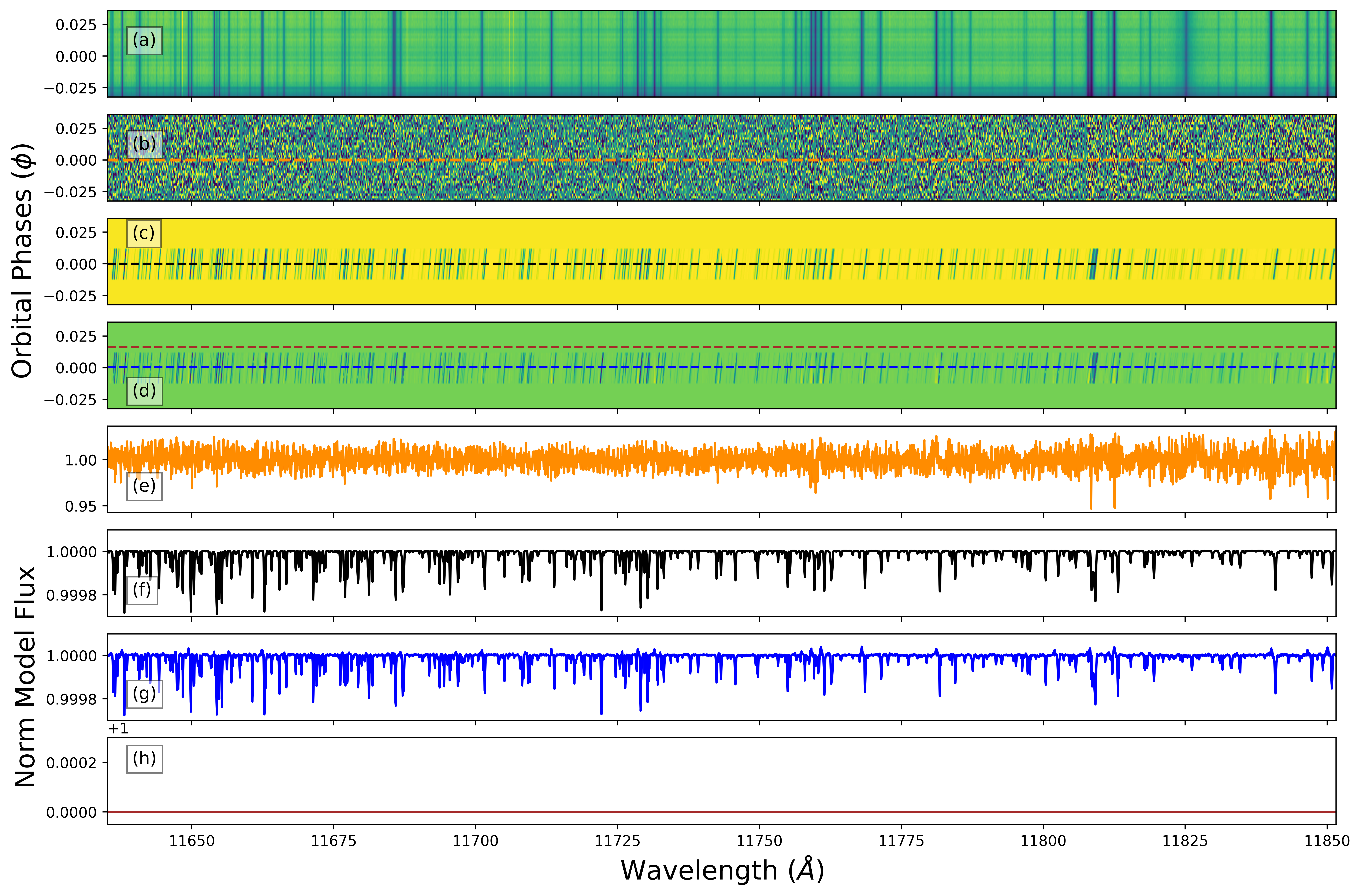}
    \caption{Same as Figure \ref{fig:steps} but with the detrending process done in the wavelength domain instead. We don't see any out-of-transit artefacts in Panels (d) and (h) in this case.}
    \label{fig:steps_wavdom}
\end{figure*}


\bsp	
\label{lastpage}
\end{document}